# Local Strain-Dependent Anisotropy in Fibrous Networks

Yoni Koren[a,b], Shahar Goren[a,b,c], Oren Tchaicheeyan[a,b] and Ayelet Lesman[a,b]

[a]*School of Mechanical Engineering, Faculty of Engineering, Tel Aviv University, Tel Aviv 69978, Israel*
[b]*Center for Physics and Chemistry of Living Systems, Tel Aviv University, Tel Aviv 69978, Israel*
[c]*School of Chemistry, Faculty of Exact Sciences, Tel Aviv University, Tel Aviv 69978, Israel*

**Abstract**

Cells in connective tissues reside within the extracellular matrix (ECM), which consists of a fibrous mesh that exhibits non-linear strain-stiffening behavior, driven by a transition from bending-to-stretching-dominated deformation. While bulk rheology captures macroscopic mechanical properties, cells actively sense and respond to local microscale heterogeneities and stiffness anisotropy in their environment. Characterizing ECM micromechanics is therefore essential for understanding the mechanical cues experienced by cells. This study quantifies local stiffness anisotropy in stretched fibrous gels by combining experimental and numerical approaches. Experimentally, we utilized optical tweezers microrheology to measure local stiffness in fibrin gels subjected to uniaxial stretch. The gels demonstrated gradual local stiffening along both the tensile and perpendicular axes, with a more profound increase along the tensile axis, resulting in local anisotropy. To investigate the physical parameters driving this phenomenon, we developed a 3D finite element model of a discrete random fiber network, successfully replicating the experimental local stiffening and anisotropy. Numerical analysis further revealed that within the sub-isostatic region, both fiber thickness and network connectivity strongly influence local anisotropy: slender fibers and higher connectivity amplify the anisotropy by up to an order of magnitude. This contributes to the formation of a highly anisotropic local environment, thereby playing a significant role in directing mechanically driven biological processes, such as cell migration and durotaxis. Our simulations also indicate that local micromechanical responses may differ from the material's global stiffening behaviors, highlighting the need for characterization at the microscopic scale.

* Corresponding author E-mail address: ayeletlesman@tauex.tau.ac.il

## 1 Introduction

Within animal connective tissues, cells are embedded in a fibrous semi-flexible biopolymer network known as the extracellular matrix (ECM). The ECM possesses unique rheological signatures distinct from synthetic flexible polymer gels. A hallmark of these networks is their pronounced strain-stiffening behavior—a nonlinear elastic response where the differential modulus increases by orders of magnitude under stress (Storm et al., 2005; Stein et al., 2011). This phenomenon is mechanistically underpinned by a geometric transition from bending-dominated to stretching-dominated deformation modes (Licup et al., 2015; Sharma et al., 2016a; Sharma et al., 2016b; Feng et al., 2016). In the linear regime at low strains, the macroscopic network response is compliant, as deformation is accommodated primarily through the bending of individual filaments and non-affine network rearrangements. However, as the applied strain increases, filaments reorient and align with the principal axis of stress. Once the soft bending modes are exhausted, the deformation increasingly requires axial stretching of the fibers. This bending-to-stretching transition provides a critical physiological support to the cells (Chaudhuri et al., 2020), allowing tissues to remain compliant at rest while robustly resisting large deformations to maintain physical integrity, and facilitates biochemical and biomechanical signaling between cells (Pakshir et al., 2019).

While bulk rheology is very common for the mechanical characterization of gels, it is in general insensitive to local micron scale heterogeneities and there can be fundamental differences between the global and local properties of the gels, with unique features which are not captured via global measurements, such as non-affine deformation and spatial variations in the gel stiffness (Chandran and Barocas, 2006; Kotlarchyk et al., 2011; Shayegan and Forde, 2013; Jones et al., 2015; Burkel et al., 2018). It has been widely reported that cells sense and respond to local differences in substrate stiffness (Tan *et al.*, 2003; Discher et al., 2005; Janmey et al., 2020), which results in the regulation of multiple biological activities such as cell movement and migration (Lo et al., 2000), cell differentiation (Engler et al., 2006), and cancer progression (Paszek et al., 2005). Cells also sense and generate forces, that propagate over long distances in the ECM compared to the cell size, reaching tens of cell diameters away (Ma et al., 2013; Wang et al., 2014) and alter the structure and stiffness of the ECM (Winer et al., 2009; Hall et al., 2016). This is often manifested by the formation of aligned and dense fiber bands connecting neighboring cells (Kim et al., 2017; Natan et al., 2020). These bands constitute regions of high rigidity relative to other areas of the substrate, creating a mechanically anisotropic environment and consequently driving cellular processes (Vader et al., 2009; Shi et al., 2014; Ban et al., 2018; Nahum et al., 2023). Therefore, it is important to characterize the ECM mechanics at the scale of the cells.

One of the most prominent techniques for microscale characterization is optical tweezers microrheology. It utilizes a tightly focused laser beam to apply controlled forces to micron-sized particles suspended in the medium and measure their response. Optical tweezers have previously been used to measure heterogeneity of viscoelastic properties during fibrillar self-assembly development (Shayegan and Forde, 2013), cell-induced local stiffness variations in fibrous networks (Jones et al., 2015; Han et al., 2018) ,as well as stiffness heterogeneities in gels under external mechanical loads (Kotlarchyk et al., 2011). The effect of external loading on the ECM is intriguing as tissues in living organisms are constantly subjected to external stresses, and local mechanical characterization of gels in physiologically relevant conditions is lacking. In a recent study, we showed that the aligned fiber bands between cells stiffen dramatically relative to the surrounding gel, and also become substantially anisotropic in their stiffness (Goren et al., 2024). This expanded on previous works which have demonstrated that cells tend to stiffen their local environment and increase the micromechanical anisotropy in their vicinity (Kim et al., 2017; Thrivikraman et al., 2021). However, it is unclear whether these experimental findings are a phenomenon that characterizes fiber networks in general, what the physical parameters that govern the mechanics at these scales are, and whether and to what extent the anisotropy in stiffness can be tuned.

In this work, we combine optical tweezers experiments in strained gels with a numerical model. We quantify the local stiffness anisotropy in stretched fibrous gels for increasing levels of strain, both experimentally and numerically, and show a systematic increase of this anisotropy with strain. Additionally, we study the effect of physical network parameters, such as fiber thickness and network connectivity, on the local stiffness anisotropy. We study networks with low connectivities (that is, at the sub-isostatic region), as according to the concept of the central-force isostatic point, the onset of mechanical instability for a system of particles with central force interactions occurs when the average connectivity of bonds is twice the spatial dimension, i.e., $<z>=2d$, which in the case of 3D networks is equal to 6 (Maxwell, 1864). Below this point the network is considered floppy, and is expected to exhibit a transition to a rigid phase when subjected to external strain. We hypothesize that this phase transition is necessary for significant stiffening and local anisotropy in fiber networks.

Our numerical analysis reveals that below the isostatic point, both fiber thickness and network connectivity have a remarkable effect on the stiffening of the network - either globally or locally. The smaller the fiber thickness – the higher the anisotropic ratio becomes, as well as the stiffness heterogeneity throughout the gel. Increasing the average connectivity of the network further increases the anisotropy and heterogeneity. Our results show that the observed local stiffening and anisotropy is correctly predicted by our model, which relies on a transition from a bending-dominated low strain regime to stretching-dominated high strain regime. This strengthens our understanding of the micromechanics of biopolymer networks, and opens ways for better modeling of mechanically driven biological processes.

## 2. Methods

### 2.1 Biological experiments

#### 2.1.1 Gel preparation

Fibrin gels were prepared by mixing human fibrinogen (Evicel Biopharmaceuticals) at a final concentration of 0.5-1 mg/ml with human Thrombin (Evicel Biopharmaceuticals) at a final concentration of 1 U/ml. The fibrinogen was fluorescently labeled with Alexa Fluor 546, succinimidyl ester (Invitrogen), as described in a previous study (Roitblat Riba *et al.*, 2019). The thrombin solution was supplemented with 0.01% wt of 0.75 µm fluorescent beads (Fluoresbrite ®YG Carboxylate Microspheres, Polysciences) to serve as tracers, and 0.008% wt of 3.5 μm fluorescent beads (Fluorescent Sky Blue Particles, Spherotech) to be used for microrheology. Fibrinogen and thrombin cast and mixed within a circular cutout in a silicone rubber strip and covered from top and bottom with thin PDMS layers, as described previously (Goren *et al.*, 2024). The gels were then set for polymerization in 37 C◦ for 2 h. Microrheology experiments were conducted within 24 h after casting.

2.1.2 Local stiffness measurements using optical tweezers

The gels, encapsulated within the silicone strips, were stretched in steps of 6mm elongation using a stretching apparatus described previously (Goren et al., 2024). Microrheology experiments were conducted using C-Trap ®confocal fluorescence optical tweezers setup (LUMICKS) made of an inverted microscope based on a 60X water-immersion objective (NA 1.2) together with a condenser top lens. The optical trap is generated by a 10 W 1064-nm laser. The trap can be steered using a piezo mirror with a feedback mechanism to determine its accurate position at 78 KHz. The samples were illuminated by an 850-nm LED and imaged in transmission onto a metal-oxide semiconductor (CMOS) camera at a frame rate of 125 Hz for calibration and 30 Hz for microrheology experiments. The bead position was tracked by the camera using automatic template matching, as well as by a back-focal plane position-sensitive diode that detects the deflection of the laser beam.

The local strains in the gel were estimated using an elongation-strain calibration curve obtained in our previous work (Goren et al., 2024). For each strain, the local stiffness was measured in parallel and perpendicular to the stretch axis, for 15-25 individual beads. The measurement was performed by oscillating the optical trap in the chosen axis for 5 seconds at a frequency of 2 Hz and amplitude of 1 μm, and tracking the bead's response. The estimated stiffness was obtained from the amplitude and phase of the bead's response, given by (Fischer and Berg-Sørensen, 2007):

$$G = Re\left\{\frac{\kappa}{6\pi a}\left(\frac{A_{trap}}{A_{bead}}\exp(i\Delta\phi) - 1\right)\right\} \quad (1)$$

where $\kappa$ is the trap stiffness, $a$ is the bead's radius, $A_{trap}$ and $A_{bead}$ are the oscillation amplitudes of the trap and bead respectively, and $\Delta\phi$ is the phase difference between the trap and bead oscillations.

**2.2 Computational modeling**

2.2.1 Model geometry

To study how the global properties of the gel may correlate to its local behavior, and to gain a deeper understanding of the key parameters that determine the degree of local anisotropy in the gel, we developed a 3D finite element (FE) model of a discrete fiber network. The geometry of the model was constructed in a similar manner to that presented in a previous publication (Sopher et al., 2023), using MATLAB. A cubic lattice of 19×19×19 face-centered-cubic (FCC) repeating units was initially generated to model the network in its organized state, with each unit consisting of segments that simulate the fibers (Supplementary Figure S2A). These units were connected by nodes, which act as crosslinks. Thus, each node in the lattice (not including the nodes at the boundaries) was initially connected to 12 other nodes, resulting in a network connectivity (i.e., the number of segments meeting at each node) of 12. To match the connectivity value to biological biopolymer networks - and specifically fibrin gels, which naturally have lower average connectivities (often ranging from $z$ = 3 to $z$ = 4, (Carlisle et al., 2010; Jansen et al., 2018)), segments were randomly eliminated from the network until a desired average connectivity value was achieved, which in our case was 3.7 or 5 (the probability weight for all nodal connectivities is shown in Supplementary Figure S2B). To make the fibers isotropically oriented, the locations of all nodes were adjusted by displacing each node to a random location contained within a spherical region of a radius equal to the fiber length. This way, we designed the network to have an approximately isotropic distribution of fiber orientations (except for a slight directional preference that still exists as a result of the original geometry of the network) and a variety of fiber lengths. Each fiber had a circular cross section $A = \pi r^2$ with a constant radius $r$ of either 50 nm, 150 nm, or 250 nm, and a length $l$ ranged from 1 μm to 17 μm according to a random distribution centered at 7 μm, typical to individual fibrin fibers (Collet et al., 2005; Beroz et al., 2017; Pancaldi et al., 2022) (Supplementary Figure S2C).

### 2.2.2 Fiber mechanical properties and energy contributions

Each fiber was modeled using two linear Timoshenko beam elements with linear elastic behavior. Note that dividing the fibers into a larger number of elements did not change the results substantially, however it resulted in a significant increase in calculation time, so the simulations throughout this work were run with fibers composed of two elements. The definition of at least two linear elements for each fiber is critical for allowing the fiber to undergo bending. The nodes connecting the fibers were assumed to be welded, enabling the transmission of both forces and bending moments. Thus, the fibers can undergo stretching, compression, bending, twisting, and rotation, depending on the load acting on them. Each fiber had an elastic modulus $E_f = 10\ MPa$, which is a typical value for crosslinked individual fibrin fibers (Collet et al., 2005; Litvinov and Weisel, 2017). The second moment of inertia of each fiber is $I = \pi r^4/4$, which is suitable for beams with a circular cross-section. If we denote the fibers axial stiffness as $\mu = EA$ and the fibers bending stiffness as $\kappa = EI$ (Landau et al., 1986), and define the dimensionless bending stiffness as $\tilde{\kappa} = \kappa/\mu l^2$, we find that $\tilde{\kappa} = r^2/4l^2$. Here, based on the range of fiber lengths and radii we used, we get a range of $\tilde{\kappa}$ in the orders of $10^{-6}$ to $10^{-3}$, with an average value of $\sim 10^{-4}$, a typical value for individual biopolymer fibers (Licup et al., 2015; Sarkar and Notbohm, 2022). The total elastic energy of the network, $U$, can be achieved by summing each of the contributions, that is, the stretching energy $U_s$, the bending energy $U_b$, the torsional energy $U_t$, and the shear energy $U_{shear}$. Nevertheless, the fibers in our networks can be considered as slender beams (i.e., $l/r > \sim 20$), so the shear energy can be neglected since shear deformation effects become negligible. Therefore, we can assume that $U \cong U_s + U_b + U_t$. Each of the energy contributions can be evaluated as a sum of the energies of all individual fibers in the network (total of $n$ fibers, or $2n$ elements), and were calculated as follows:

$$U_s = \sum_{i=0}^{2n} \int \frac{\mu}{2} \left(\frac{\partial u_i}{\partial s}\right)^2 ds \quad (2)$$

$$U_b = \sum_{i=0}^{2n} \int \frac{\kappa}{2} \left(\frac{\partial \Psi_i}{\partial s}\right)^2 ds \quad (3)$$

$$U_t = \sum_{i=0}^{2n} \int \frac{GJ}{2} \left(\frac{\partial v_i}{\partial s}\right)^2 ds, \quad (4)$$

where $G$ is the shear modulus of the $i$th element, $J$ is the second polar moment of area of the element, $\partial u/\partial s$ is the axial strain, $\partial \Psi/\partial s$ is the rotation of the plane perpendicular to the axis along the fiber contour (i.e., curvature change), $\partial v/\partial s$ is the rotation of the fiber cross-section, and $s$ is the axis along the fiber contour. The shear modulus of each element is obtained by $G = \frac{E}{2(1+\nu)}$, and the second polar moment of area is $J = 2I = \pi r^4/2$.

To obtain an approximate value of the axial, logarithmic (true) strain along the element, strains were calculated at the centroid of each element. This way we can consider these values as constants along the elements, and eq. (2) is simplified to the following:

$$U_s = \frac{\mu}{2} \sum_{i=0}^{2n} l_i {\varepsilon_i}^2, \quad (5)$$

where $\varepsilon_i$ is the logarithmic strain at the centroid of the $i$th element, $l_i$ is the length of the $i$th element, and $2n$ is the number of elements (segments) in the network.

The curvature change for each element was calculated by dividing the difference of rotational displacements of the two nodes connected to the element by the element length. Eq. (3) is then simplified to the following:

$$U_b = \frac{\kappa}{2} \sum_{i=0}^{2n} l_i \left(\frac{ur_{2,i} - ur_{1,i}}{l_i}\right)^2, \quad (6)$$

where $ur_{1,i}$ and $ur_{2,i}$ are the components of rotational displacements of the $i$th element two nodes that belong to the plane perpendicular to $s$ axis.

The rotation of the fiber cross-section is obtained by dividing the difference of rotational displacement components along $s$ axis of the two nodes connected to the element by the element length, so Eq. (4) is simplified to the following:

$$U_t = \frac{1}{2} \sum_{i=0}^{2n} GJ l_i \left(\frac{ur_{s2,i} - ur_{s1,i}}{l_i}\right)^2, \quad (7)$$

where $ur_{s1,i}$ and $ur_{s2,i}$ are the rotational displacement components along $s$ axis of the two nodes of the $i$th element.

To characterize the network during stretching, alongside the energetic contributions, it is also beneficial to analyze the collective reorientation of the fibers. We employed the 3D nematic order parameter, that is defined as

$$S = \frac{1}{2}\langle 3cos^2\theta - 1\rangle, \tag{8}$$

where $\theta \in \left[0, \frac{\pi}{2}\right]$ is the angle between an individual fiber's axis and the stretching direction (y).

2.2.3 Simulations setup and boundary conditions

Two different scenarios were considered (Figure 2D): (i) Uniaxial tensile simulations, which aim to characterize the global mechanics of the network and to study the relation between the transition of mechanical regimes (bending-dominated to stretching-dominated) to the local mechanical anisotropy emerged in the network. This was done by constraining the nodes at the bottom face of the network in the vertical direction (y – the direction of stretch), and subjecting the nodes at the top face to uniaxial tension by applying vertical maximum displacement of 60 μm to these nodes. Each side of the cubic lattice was initially 95 μm long, so the applied displacement is equivalent to an engineering strain of ~63%. The lateral faces had zero traction, so that the network was able to undergo free lateral contraction while stretching vertically. The simulation step duration was 300 seconds, so the stretching velocity was one-third of a μm per second (a simulation that is slow enough and approximately quasi-static to obtain time-independent network behavior). Note that stretching velocity lower than one-third of a μm per second resulted in a very similar response of the network, allowing us to verify that the network was indeed stretched in a time-independent manner. (ii) Local measurements of stiffness in both stretching and transverse directions. The purpose was to assess the local mechanical anisotropy in the network. This was done by stretching the network to a certain strain for 12.5 seconds in the same manner as in the uniaxial tensile simulations, followed by a 5 second pause to allow the kinetic energy to dissipate, and then fixing the nodes at the top face and the side faces and applying a 0.2-μm displacement (i.e., point displacement) to a single node in the middle of the network, for another 12 seconds. Pulling the node at such a slow speed and by such a small amount ensures a negligible change in the state of the network, and yet allows for measurement of the network's reaction force to the displacement. In each level of stretch, the point displacement was applied along the stretching direction (y) and the transverse direction (x) separately, and the resulting reaction force component along this direction was reported for the stiffness calculation. Since the force-displacement relation was linear at such small displacements range, we could simply refer to the slope of the curve as the local stiffness in each direction. In cases where the relation between the resulting force and the applied displacement was not entirely linear (following a non-linear network response caused at high tensile strains—typically above the critical strain), the local stiffness was determined by fitting a linear curve. Consequently, the *stiffness fold change* is obtained by dividing the stiffness in each level of stretch by the **initial** stiffness measured at that node.

2.2.4 Numerical methods

We used the commercial FE software Abaqus/CAE 2023 (Dassault Systemes Simulia, Johnston, RI) to simulate the mechanics of the networks. As mentioned before, we used two two-node linear beam elements (type B31) to model the fibers, as it is crucial for bending. Moreover, it allows for better numerical convergence of the simulation at high levels of strain, compared to truss elements that can only transmit axial forces and therefore might simulate the network mechanics under external loading less realistically. The large displacements in the network, occurring at high external strains, introduced a high level of geometric nonlinearity that precluded the use of an implicit static solver. Therefore, the software's explicit dynamic solver was utilized with sufficient damping and time to attain a quasi-static steady state at each displacement increment (the quasi-static steady state was defined to be reached when the model's kinetic energy dissipates to 0.01% of its strain energy). We chose $\alpha = 0.1$ and $\beta = 10^{-5}$ as the Rayleigh damping coefficients, using smooth amplitude time steps for both uniaxial stretching and point displacement. These settings were chosen to produce the fastest convergence to steady state, while ensuring that the simulation outcome remains nearly the same as the static solution at small displacements, and dynamic effects remain insignificant. The network was composed of approximately 12,000 nodes and 14,400 elements (for connectivity of 3.7) or 18,000 (for connectivity of 5); this allowed, on the one hand, a reasonable running time for the simulations, and on the other hand, a sufficiently large network to make sure that the displacements near the displaced node in the middle of the network will not be affected by the boundaries of the network. Local measurements of stiffness were done for 14 different nodes located in the vicinity of the center of the network. For each of the nodes in this sample, the stiffnesses along the x and y axes were measured in increments of 2.5% external strain, using a Python script that generated the simulations for each strain, and another script that was used to analyze the results.

## 3 Results

### Microrheology of stretched fibrin gel demonstrates local stiffness anisotropy

To study how the micromechanics of fibrous gels are affected by an applied strain, we used a 3D-printed apparatus to stretch fibrin gel samples (0.5 and 1 mg/ml) in a controlled manner. This apparatus was designed so that it could be mounted on the microscope stage and simultaneously allow microrheology experiments to be performed. The gels, which were embedded in a cavity in the center of silicone strips attached to the device, were stretched in multiple steps. The strain at each step was obtained from using a previously obtained calibration (Goren et al., 2024). At each stretching step, microrheology measurements of the local stiffness were performed. Figure 1A-C shows fluorescence images of a 1 mg/mL fibrin gel during stretching, where realignment of the fibers in the direction of stretching can be seen. The local stiffness was measured according to according to Eq. (1) by oscillating a fluorescent bead (i.e., optical trap) embedded in the gel in both parallel and perpendicular to the stretch axis, and then monitoring its response in both directions (see Methods). For each gel concentration, measurements were performed on three samples, with the local stiffness measured for 15-25 beads in each step. The change in average local stiffness relative to the initial local stiffness (stiffness fold change) along the axis perpendicular to the stretch (x-axis) and in stiffness along the axis parallel to the stretch (y-axis) are shown in Figure 1D. Gradual local stiffening occurs along both the tensile and perpendicular axes, with a more significant stiffening occurring along the tensile axis, for both gel concentrations. Interestingly, the lower concentration gel (red curves) demonstrated higher stiffening. For example, at 22% strain, 0.5 mg/ml gels stiffened 10.5 ± 1.5-fold along the tensile axis and 4.5 ± 1-fold along the perpendicular axis, while 1 mg/ml gels stiffened only 5.5 ± 0.5-fold along the tensile axis and 2.2 ± 0.1-fold along the perpendicular axis. The local anisotropy, which was evaluated by dividing the stiffness fold change along the tensile axis by the perpendicular axis for each individual bead, indicates that the gels become locally anisotropic under tension (Figure 1E). At both concentrations, the level of anisotropy varied between similar values of up to 3 (at 33% strain). These results are in accordance with our previous experimental results (Goren et al., 2024). Data for individual gels is shown in Supplementary Figure S1.

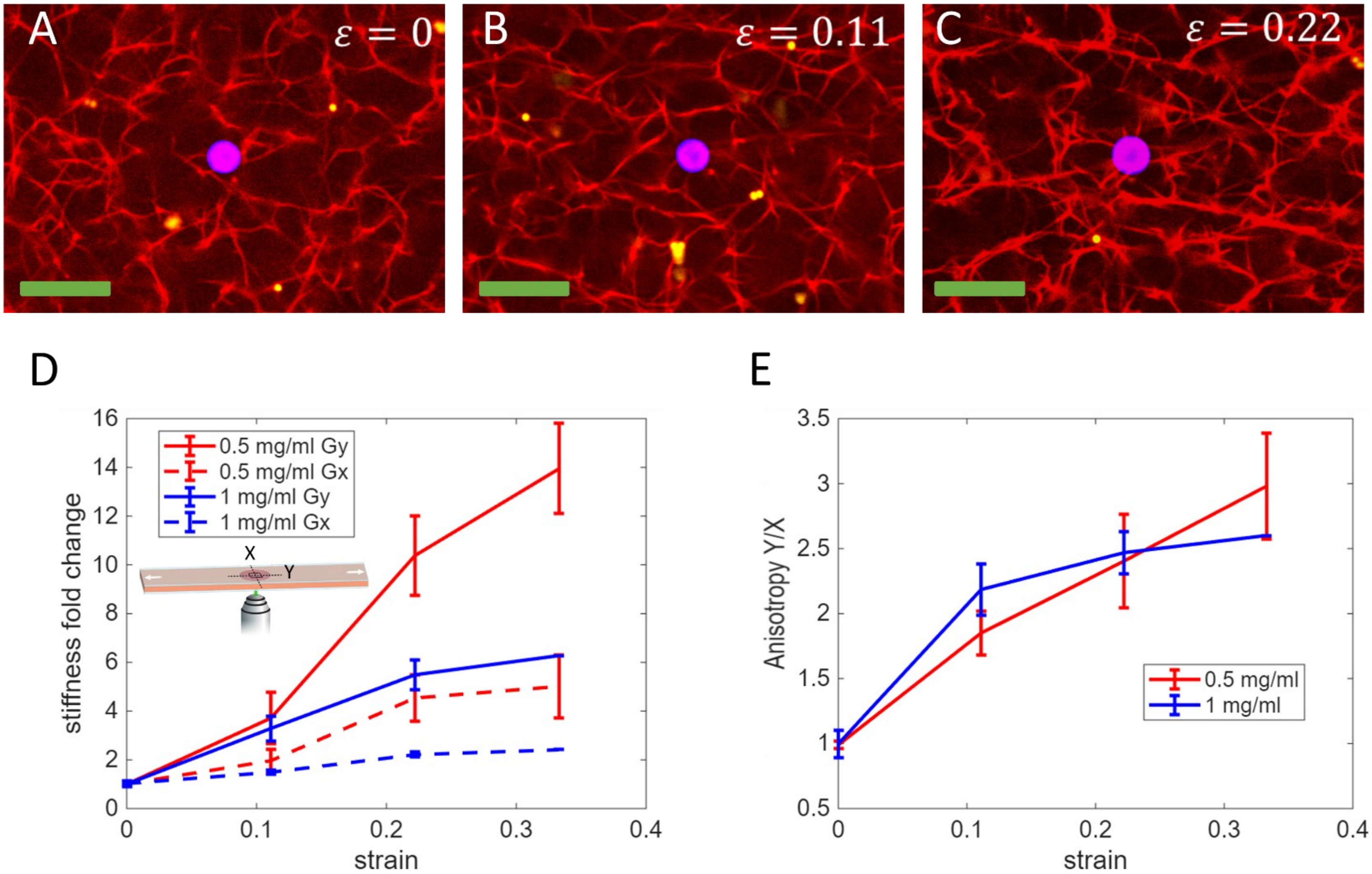

Fig. 1. Microrheology of stretched fibrin gels. (A-C) Confocal images of a 1 mg/ml fibrin gel under 0, 11, and 22 % strain. The fibers are shown in red and a reporter bead in purple. Scale bars – 10 μm. (D,E) Fold-change in local stiffness (D) and Y/X anisotropy (E) at 2 Hz in the x and y directions with increasing strain, for 0.5 and 1 mg/ml fibrin. Error bars are mean ± standard error of three different samples with 15-25 measurements per sample. Data for individual gels is shown in Supplementary Figure S1. Inset in panel (D) shows an illustration of the stretching apparatus.

**A 3D model of discrete random fiber network**

To understand the physical parameters that control local stiffening and anisotropy, we complement our experimental analysis with a computational framework using finite-element (FE) model. A discrete 3D fiber network was constructed. The network geometry (100×100×100 μm) is based on a face-centered cubic (FCC) lattice, randomly diluted down to a desired connectivity number (3.7 or 5.0) (Beroz et al., 2017; Sopher et al., 2023). Whereas $\langle z \rangle = 3.7$ is a typical value for biological gels, we also study networks of $\langle z \rangle = 5.0$. Each network node was randomly shifted in the stress-free state to obtain a variety of fiber lengths orientations (Methods). Lengths, diameters, orientations, and densities of the fibers in the resulting simulated network qualitatively capture the fibrin gels used in the experiments (Figure 2A, Supplementary Figure S2). The fibers had diameter $D_f$ of 100, 300, or 500 nm. Each fiber was modeled as a linear elastic beam composed of two straight segments, and can thus stretch, compress, and bend due to an applied local force (Figure 2B).

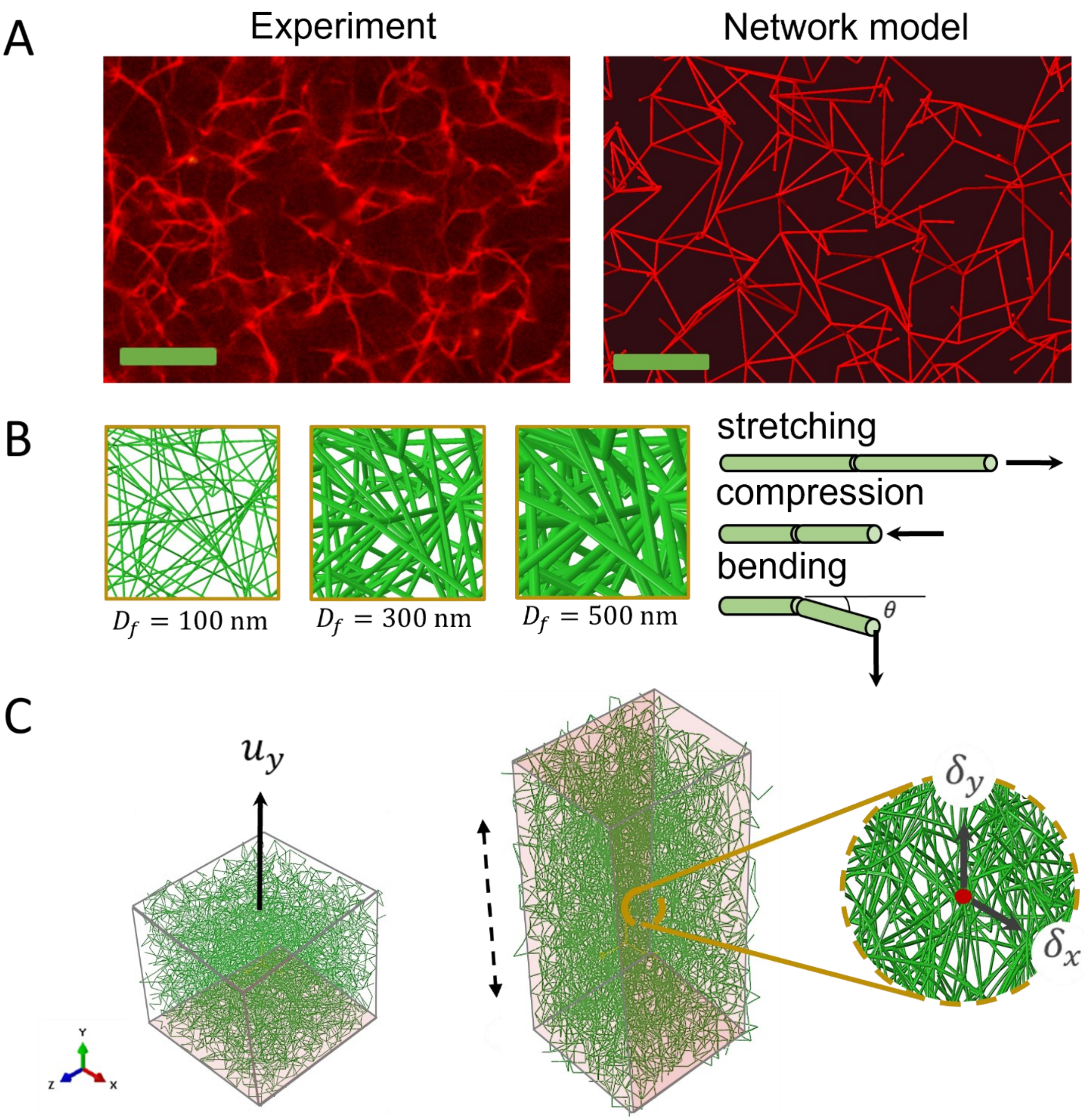

Fig. 2. Fibrous network model. (A) Confocal image of a fluorescently-labeled 1 mg/ml fibrin gel (left), compared with a thin slice of the 3D fibrous network model ($<z> = 3.7$, $Df = 300$nm; right). Scale bars – 10 µm. (B) The network is composed of fibers with a diameter of 100, 300, or 500 nm, which can be stretched, compressed, or bent. The fibers consist of two segments hinged together to allow bending. (C) The network is stretched by applying an upward displacement to the nodes on the top surface, while the nodes on the bottom surface remain fixed. To measure local stiffness in the stretched network, a point displacement is applied at the center, in either the x- or y-direction ($\delta_x$ and $\delta_y$, respectively).

**External tension causes global stiffening and a transition from bending to stretching**

We first studied the global response of the network to external stretch for the two connectivity levels and the three fiber thicknesses mentioned above. The nodes at the bottom of the network were fixed, whereas the nodes at the top were pulled upwards (in the positive y direction; Figure 2C). As expected, in all cases the network stiffens nonlinearly under tension, while the thicker the fiber, the steeper the increase in stress with external strain (Figure 3A). Increasing <*z*> (i.e., the network average connectivity) from 3.7 to 5.0 resulted in an even more dramatic increase in stress, and the onset of nonlinear stiffening at an earlier stage. As individual fibers are linear-elastic, this stiffening can be explained by the fibrous nature of the network and the transition from a bending-dominated to a stretching-dominated state. The differential modulus of the network, $K$, is approximately constant until the point where the nonlinear stiffening occurs (Figure 3B-C). The network is initially dominated by bending and then a transition toward a stretching-dominated phase is observed where the stretching energy exceeds the bending energy ($\frac{U_s}{U_b} > 1$; Figure 3D, Supplementary Figure S3). The thinner the fibers, the more extreme $\frac{U_s}{U_b}$ becomes (<<1 in small strains and >>1 in large strains). We defined the transition from bending to stretching as $\varepsilon_{cr} = \frac{U_s}{U_b} = 1$. Figure 3E shows that $\varepsilon_{cr} =$ 35%-40% for the lower connectivity networks and $\varepsilon_{cr} =$ 25%-30% for the higher connectivity networks. $\varepsilon_{cr}$ occurs slightly earlier for thicker fibers with the exception of the low connectivity network with $D_f = 100\ nm$ (Figure 3D and 3E). Generally, the nonlinear stiffening curves (Fig. 3B) follow qualitatively the trends of the bending to stretching energy curves (Fig. 3D). Reorientation of fibers during stretching shows that fiber alignment occurs gradually throughout the entire stretch (Xu et al., 2019) - in parallel with the bending phase and subsequent stretching (Supplementary Figure S5). In conclusion, the 3D fiber model successfully captures the characteristic nonlinear stiffening behavior, the transition from bending-dominated to stretching-dominated mechanics, and fiber reorientation under deformation. These mechanical responses are consistent with those reported in previously established models (Sharma et al., 2016a; Feng et al., 2016; Licup et al., 2016, Zakharov et al., 2024). The model is therefore well positioned to investigate local stiffening effects induced by external stretch.

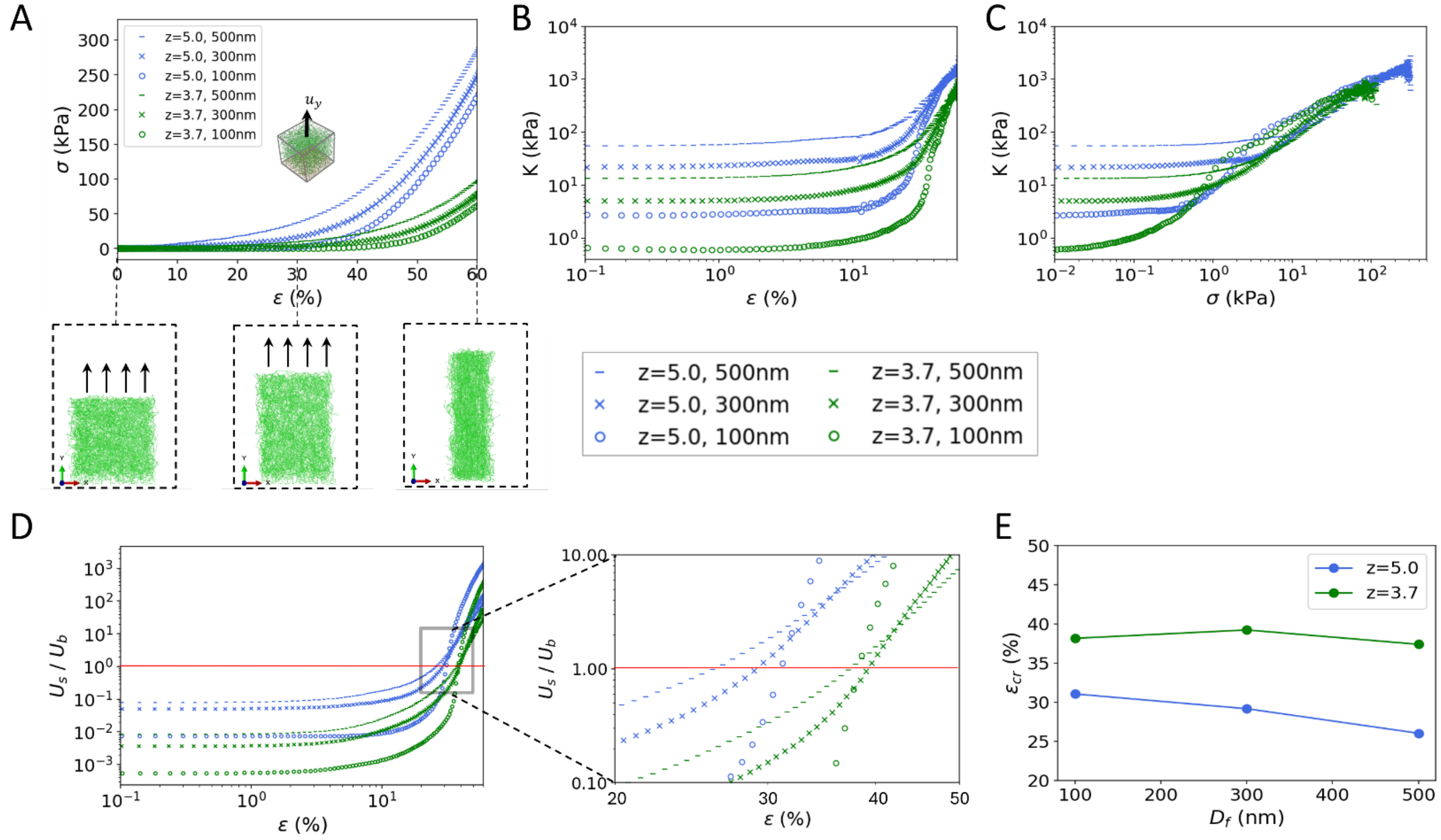

Fig. 3. Global stiffening and transition from bending to stretching: validating the mechanics of fiber networks. Two different average network connectivities are considered: <z> = 3.7 (green) and <z> = 5.0 (blue). (A) Stress vs. strain. The insets show a side view of the low connectivity, 300 nm fiber network at 0%, 30%, and 60% strain, for demonstration. (B) Differential modulus vs. strain. While the modulus depends on the fiber thickness at low strains, the onset of stiffening and convergence of the curves depend on the connectivity, as they occur earlier for higher connectivity. (C) Differential modulus vs. stress. The modulus increases with increasing fiber thickness and connectivity. (D) Ratio of stretching energy to bending energy vs. strain. In all cases, a transition from bending to stretching is shown. The red line at $Us/Ub = 1$ denotes the bending-to-stretching transition point. A zoomed in view of the bending-to-stretching transition region is shown. (E) Bending-to-stretching transition strain $\varepsilon_{cr}$ vs. fiber thickness. While fiber thickness has a minor effect on the transition strain, the effect of connectivity is more significant.

## External stretch induces local stiffening and anisotropy both in experiments and simulations

We next used the 3D fiber model to explore local stiffening and anisotropy observed experimentally. At increasing steps of stretch, we applied a point displacement to a single node in the center of the network in X and Y directions (Figure 2D) – and the local stiffnesses in both directions were obtained (Methods, and Supplementary Figure S6). Since the model is fibrous (discrete) and not continuous, the local stiffness is expected to vary depending on local geometry. Therefore, we averaged the results over a group of nodes in the center. As in the experimental system, the network stiffens locally under tension in both directions (Figure 4A-B; Supplementary Figures S7-9). Stiffening along the tensile (y) axis is greater, reaching factor of >10 for 45% strain (Figure 4A), and stiffening along the perpendicular axis (x) remains moderate, up to about a factor of 4 (Figure 4B). The anisotropy (y stiffness/ x stiffness) remains below 4 up to about 30% strain, consistent with our experimental results (Figure 1E). A high anisotropy value (>10) is obtained for a large strain of 45% for the thinnest fibers, which is beyond the level we could apply in the experimental system. This trend can also be seen in Figure 4D, showing the anisotropy and its heterogeneity below and above $\varepsilon_{cr}$. Our findings are consistent across the experimental and simulation results: both a lower gel concentration (which is equivalent to a higher slenderness of the fibers, (Sharma et al., 2016a)) in the experiments and a reduced fiber thickness in the simulations led to a higher local stiffening effect (Figure 1D, Figure 4A-B).

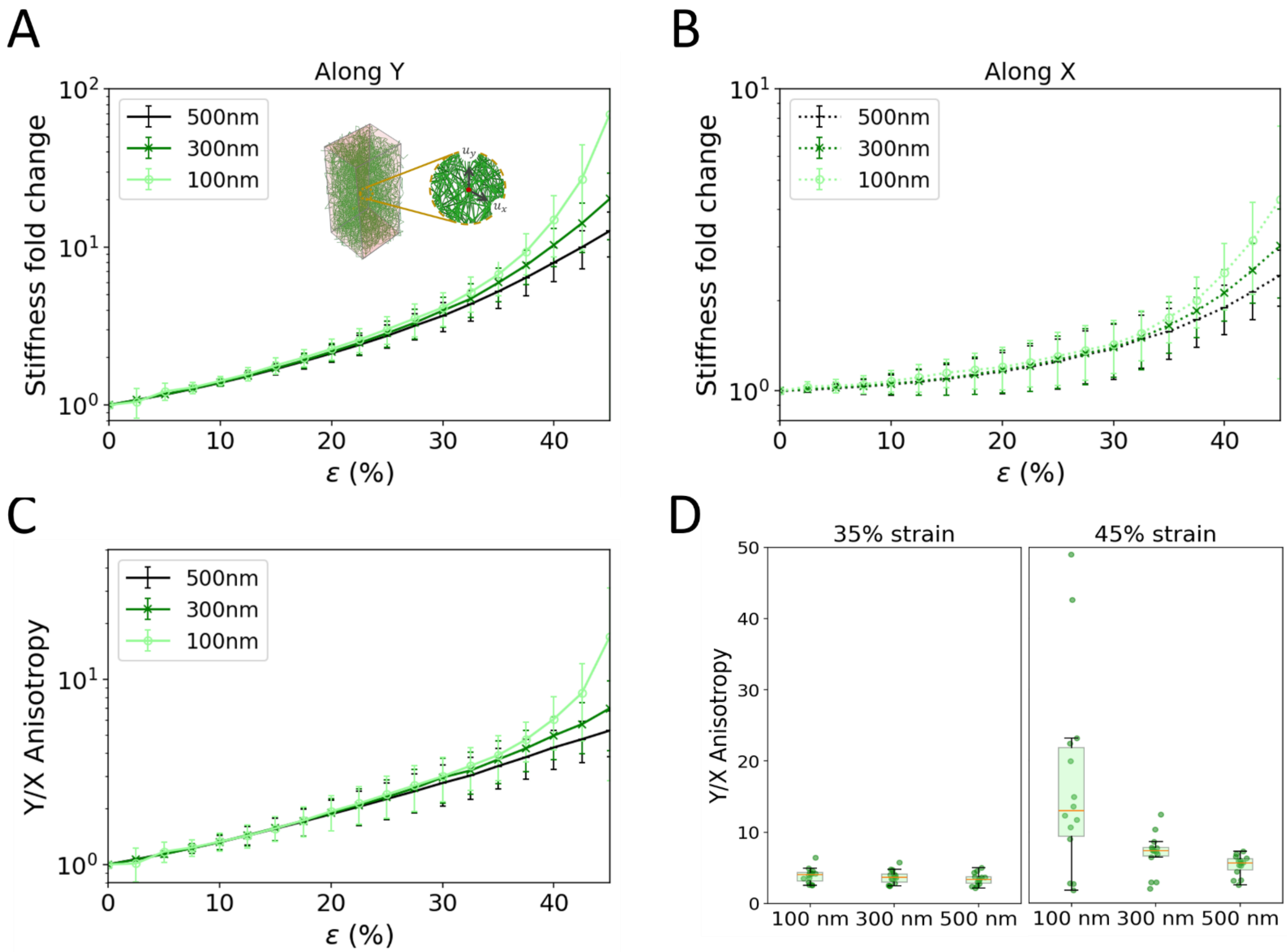

Fig. 4. Local stiffening effect in network connectivity of <*z*> = 3.7. Decreasing fiber thickness results in higher local stiffness and anisotropy. (A-B) Fold-change in the local stiffness in the y (A) and x (B) directions with increasing strain. (C) Y/X anisotropy with increasing strain. Error bars are mean ± standard error of 14 nodes. Data for each node is shown in Supplementary Figures S7-9. (D) Distributions of the Y/X anisotropy measured at the nodes in the middle of the network before and after $\varepsilon_{cr}$, corresponding to strains of 35% and 45%.

**Effect of Network Connectivity on Local Stiffening and Anisotropy**

Network connectivity is a key parameter known to determine the mechanical properties of biological networks, but how the connectivity affects stiffening was mainly studied from a macroscopic point of view (Licup et al., 2015; Sharma et al., 2016a; Sharma et al., 2016b; Feng et al., 2016). Experimentally, it is a parameter that is very difficult to control for. We thus used our model to investigate how the connectivity level influences stiffening and anisotropy at the local scale. From a global perspective, our mechanical analysis showed that increasing $<z>$ from 3.7 to 5.0 leads to more pronounced stiffening, enhanced fiber alignment in the tensile direction, and a lower critical strain for the bending-to-stretching transition (Figure 3).

At the local scale, increasing the connectivity from 3.7 to 5.0 led to greater local stiffening along both axes, with the effect being particularly pronounced for thinner fibers (Figure 5A-C; Supplementary Figures S10-12). Local anisotropy also increased with connectivity, although to a more moderate extent (Figure 5D-F). Notably, whereas reducing fiber thickness caused a substantial increase in local stiffening and anisotropy only above the critical strain, increasing connectivity enhanced both local stiffening and anisotropy across the entire strain range. For $<z>$ = 5.0, the local anisotropy measured at 25% and 35% strain (before and after $\varepsilon_{cr}$), the greatest effect is for the thinnest fibers, and decreases with increasing the fiber thickness (Supplementary Figures S13).

Finally, we directly compared the local and global stiffening responses. Overall, local stiffening followed the same trends observed at the global scale; however, a minor but consistent gap remained between the local and global responses. Interestingly, for networks with $\langle z \rangle$ = 5.0, the local stiffening exceeded the global stiffening, whereas for networks with $\langle z \rangle$ = 3.7 the opposite trend was observed, with the global response exceeding the local response (Figure 6). These findings highlight that the local mechanical response, which is the relevant scale that cells sense, may differ from the bulk mechanical behavior of the network, emphasizing the importance of measuring local micromechanics for a more accurate characterization of network mechanics.

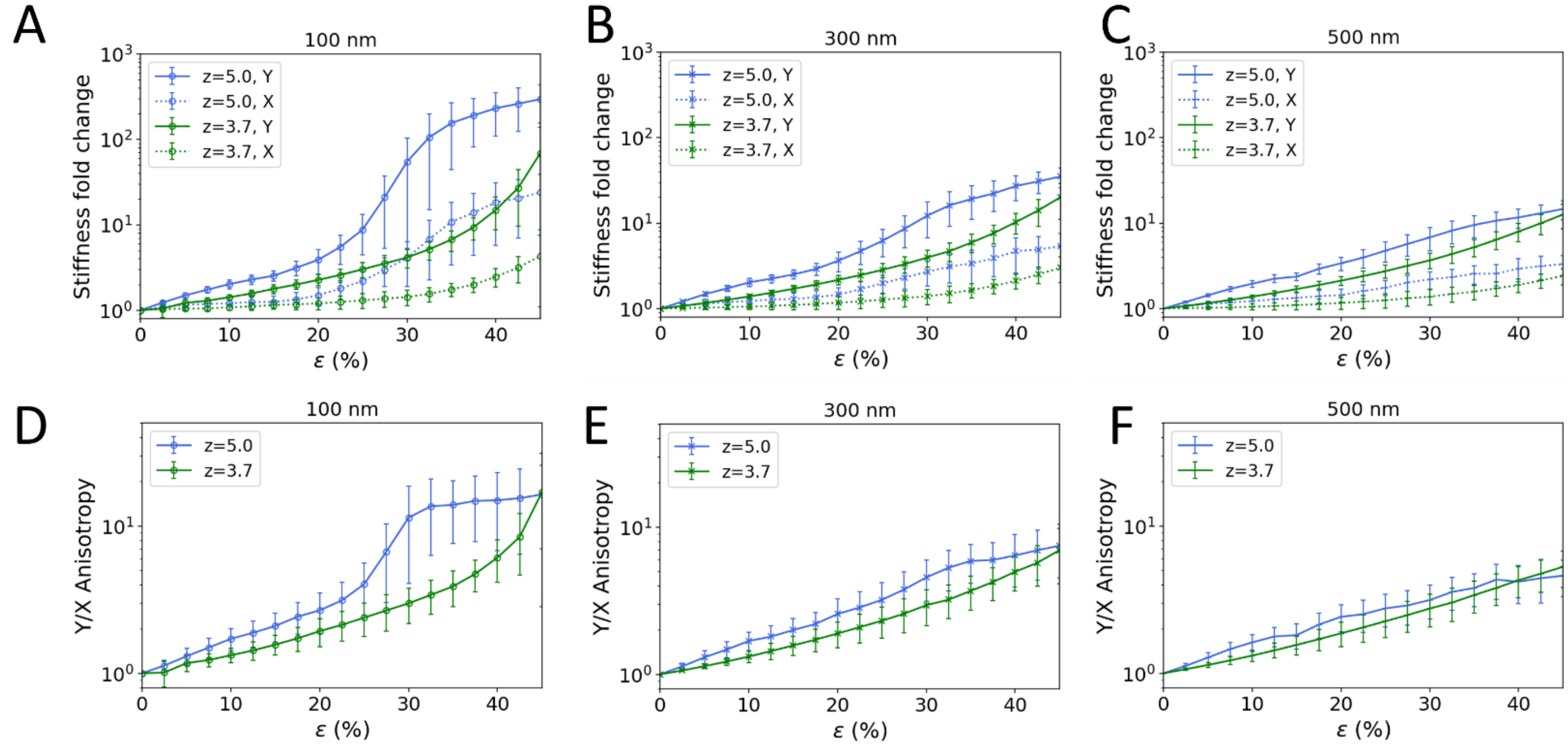

Fig. 5. Increasing connectivity leads to an increase in local stiffening and anisotropy. (A-C) Fold-change in the local stiffness for 100 nm (A), 300 nm (B), and 500 nm (C) with increasing strain. (D-F) Y/X anisotropy for 100 nm (D), 300 nm (E), and 500 nm (F) with increasing strain. Error bars are mean ± standard error of 14 nodes for <*z*> = 3.7, and 10 nodes for <*z*> = 5.0. Data for each node is shown in Supplementary Figures S10-12.

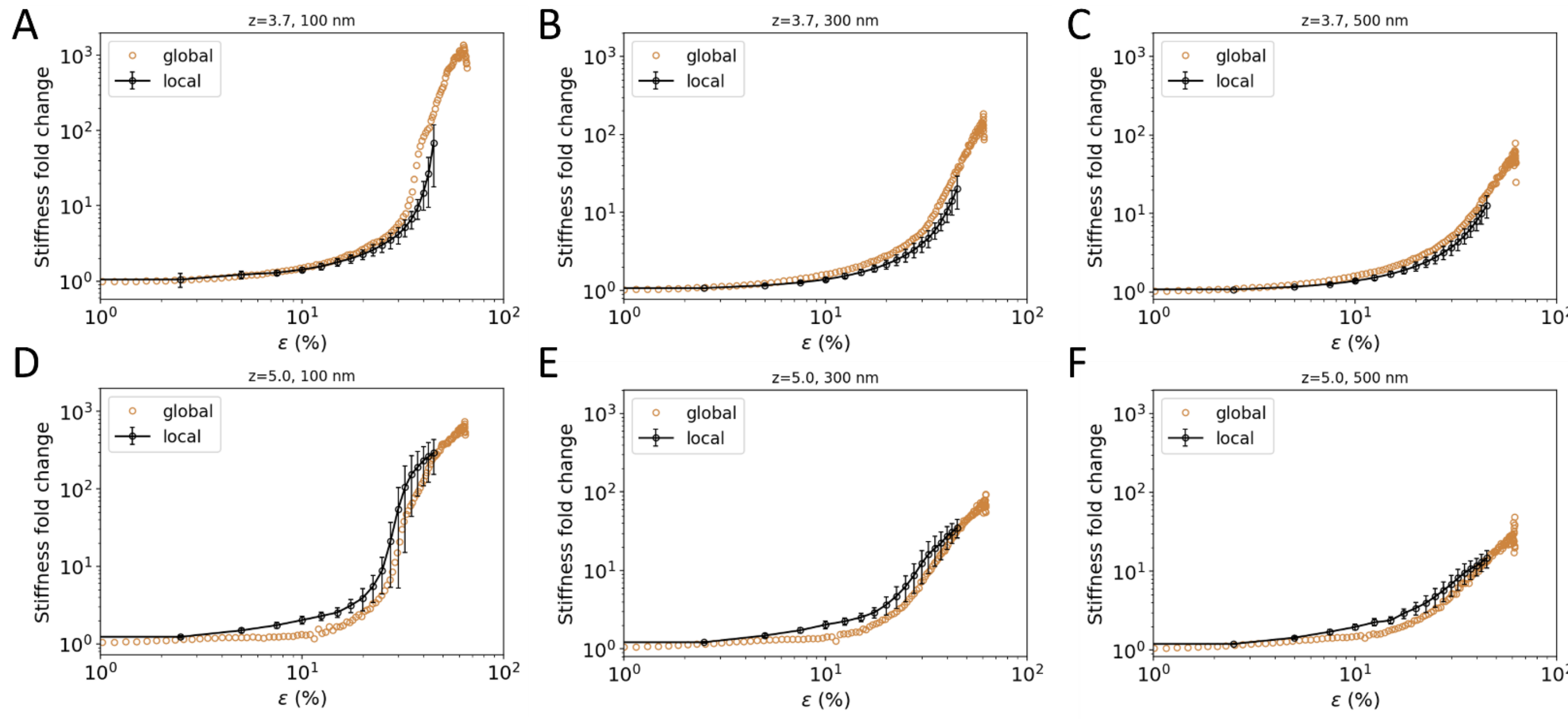

Fig. 6. Comparison between global and local stiffening in the direction of stretch (y). Stiffness fold change is shown both locally and globally for <*z*> = 3.7 (Error bars are mean ± standard error of 14 nodes) (A-C), and for <*z*> = 5.0 (Error bars are mean ± standard error of 10 nodes) (D-F).

## 4 Discussion

Previously, we investigated how fibrous gels, such as collagen and fibrin, change mechanically when subjected to physical strain. By combining a unique stretching apparatus with optical tweezers, we quantified how these gels stiffen and become anisotropic at the microscopic level (Goren et al., 2024). However, the question of which physical parameters contribute to that stiffening and increase in anisotropy, and to what extent, remained unanswered. To this end, in the present study, we developed a minimal, computational model of a fibrous network that mimics those gels and performed a comprehensive analysis - in terms of both the range of strains and physical network parameters – so that we can obtain a broader picture of the underlying mechanics in the basis of our experimental results.

The nonlinear mechanics of fibrous materials have been explained in past studies via several different mechanisms, such as the entropic nonlinearity of individual fibers (Storm et al., 2005) and fiber buckling (Rosakis et al., 2015). For collagen, evidence from rheology strongly supports the bending-to-stretching model. In this model, the small-deformation response of fiber networks is dominated by soft bending modes, which do not require the fibers to extend in length, and only as these modes gradually deplete at higher deformations, the fibers begin to stretch, leading to network stiffening. We simulated a network composed of linear beams that can bend and stretch and used it to explore the relationship between the macroscopic and microscopic mechanical properties, as well as their dependence on fiber thickness and network connectivity, and compared these results to our experiments.

In the experiments documented in the current and previous work, we observed micromechanical stiffening of a stretched network by factors of about one order of magnitude (~10 fold) in the longitudinal axis (Figure 1D), while the perpendicular axis was characterized by lower stiffening. The local anisotropy (stiffness ratio) between the longitudinal and perpendicular axes was up to about 3 (Figure 1E). It is not trivial that the network stiffens in both axes, because intuitively we would expect that the buckling of fibers oriented in the perpendicular axis (which undergo compression) might lead to softening along that axis. However, our simulations reproduce this behavior, aligning with the experimental observation that the network stiffens along both axes (Figure 4A, 4B).

Specifically, our simulations also show that below the critical strain, at which the stretching energy becomes dominant, the anisotropy is of similar values to those in our experiments (as well as in the experiments conducted in the previous work, (Goren et al., 2024)) – ranged from 1 to 3, and only beyond the critical strain, higher anisotropy is obtained.

Natural fibrous gels, such as fibrin, are typically characterized by low connectivity ranging from 3 to 4 (Jansen et al., 2018). As such, they are well below the isostatic limit for 3D networks, which is 6. Hence, they stiffen at a relatively high critical strain. In our simulations, the networks with a connectivity of 3.7 stiffen at a critical strain of ~40%, above our experimental range. To this end, our simulations are in agreement with our experimental results, assuming that our gels were below their critical strain throughout the experiment. Our simulations also go beyond the experimental strain range and reveal that they are highly anisotropic in that range.

It is important to note that there is no straight-forward relationship between gel concentration and fiber thickness in fibrin networks. In fact, networks with higher concentration were shown to have thinner fibers (Piechocka et al., 2010). The dominant factor controlling the bending-to-stretching transition, however, is the fiber slenderness (the ratio of a fiber length to its thickness), and it was shown to decrease with gel concentration following the more dramatic reduction in mesh size (Licup et al., 2015). Accordingly, higher fiber thickness in our simulations corresponds to increasing gel concentration in experiments. As a result, our simulations predict that gels with lower concentration (slender fibers) should display stronger stiffening on both axes, in agreement with the experiments (Figure 4A-B). Nevertheless, based on that same assumption, our simulations predict that different gel concentrations should not significantly affect the degree of stiffening below the critical strain. Only beyond the critical strain does the behavior of networks with different fiber thicknesses diverge. (Figure 4C). In our experiments, gels with lower concentration stiffened substantially more than gels with higher concentration. If our experiment was conducted entirely below the critical strain, our simulations predict that we should not observe this difference. In fact, it is not entirely clear from previous studies whether the stiffening trend of networks with different concentrations is indeed identical. This point requires further investigation.

Our simulations for networks with a connectivity of 5 demonstrate that these networks – being closer to the isostatic point – stiffen at a smaller critical strain (around 30%) and much more dramatically for a given strain (Figure 3). This stiffening is also apparent at the microscale (Figure 5A, 5B, 5C). These networks are more homogeneous, with their anisotropy distribution showing less outliers (Supplementary Figure S13). For collagen, previous works showed that connectivity can be tuned by controlling the temperature and gel concentration (e.g., Jansen et al., 2018), but it was still limited to values below 4. For fibrin, as far as we know, no such control has been demonstrated. Tuning the connectivity of gels can allow powerful control over their mechanics.

Looking at the stiffening of the networks at the local level, we see that relatively high heterogeneity is obtained in both axes – especially for the thinnest fibers (Supplementary Figures S7-12). The local stiffening sometimes spans two orders of magnitude across different points, as previously measured in collagen and fibrin gels (Beroz et al., 2017), demonstrating the highly heterogeneous environments in which cells function. This also may lead to differences between local stiffnesses to those measured globally, as reported previously in experimental scenarios (Kotlarchyk et al., 2011; Busenhart et al., 2025) and as we obtained here (Figure 6).

Overall, our results demonstrate that fibrous networks featuring slender filaments and connectivity near the isostatic threshold, under sufficiently high tensile strain, are primed for a more substantial increase in local anisotropy. In extracellular environments, such an increase in local anisotropy can arise not only from external stretching but also from the contractile forces exerted by cells. Contractile cells, such as fibroblasts and cancer cells, embedded within fibrous gels (e.g., fibrin or collagen) pull on the matrix fibers, creating dense and aligned fiber bundles (i.e., bands) that extend between neighboring cells (Shi et al., 2014; Kim et al., 2017; Sopher et al., 2018; Liu et al., 2020; Natan et al., 2020; Grekas et al., 2021; Doha et al., 2022; Nahum et al., 2023). We previously measured the stiffness anisotropy of these bands to be within the range of 2-3, similar to the results we obtained here for the case of uniaxial stretching. Band anisotropy was quantified as the ratio between the stiffness along the band and that in the perpendicular direction of the band (Goren et al., 2024). These anisotropic fiber bands generate stiffer, direction-dependent pathways, along which cells can migrate via durotaxis and contact guidance mechanisms (Feng et al., 2019; Szulczewski et al., 2021; Ergaz et al., 2024; Yim et al., 2026). Thus, our findings may help explain the conditions under which cells are most likely to migrate from one region to another. Moreover, anisotropy in the local mechanical environment has been shown to play a dominant role in other cell-fate decisions: in a recent study, tension anisotropy (i.e., a mechanical state where tensile forces are polarized or directional, rather than being equal in all directions) was shown to trigger the transformation of fibroblasts into myofibroblasts (Alisafaei et al., 2025).

## Supplementary Figures

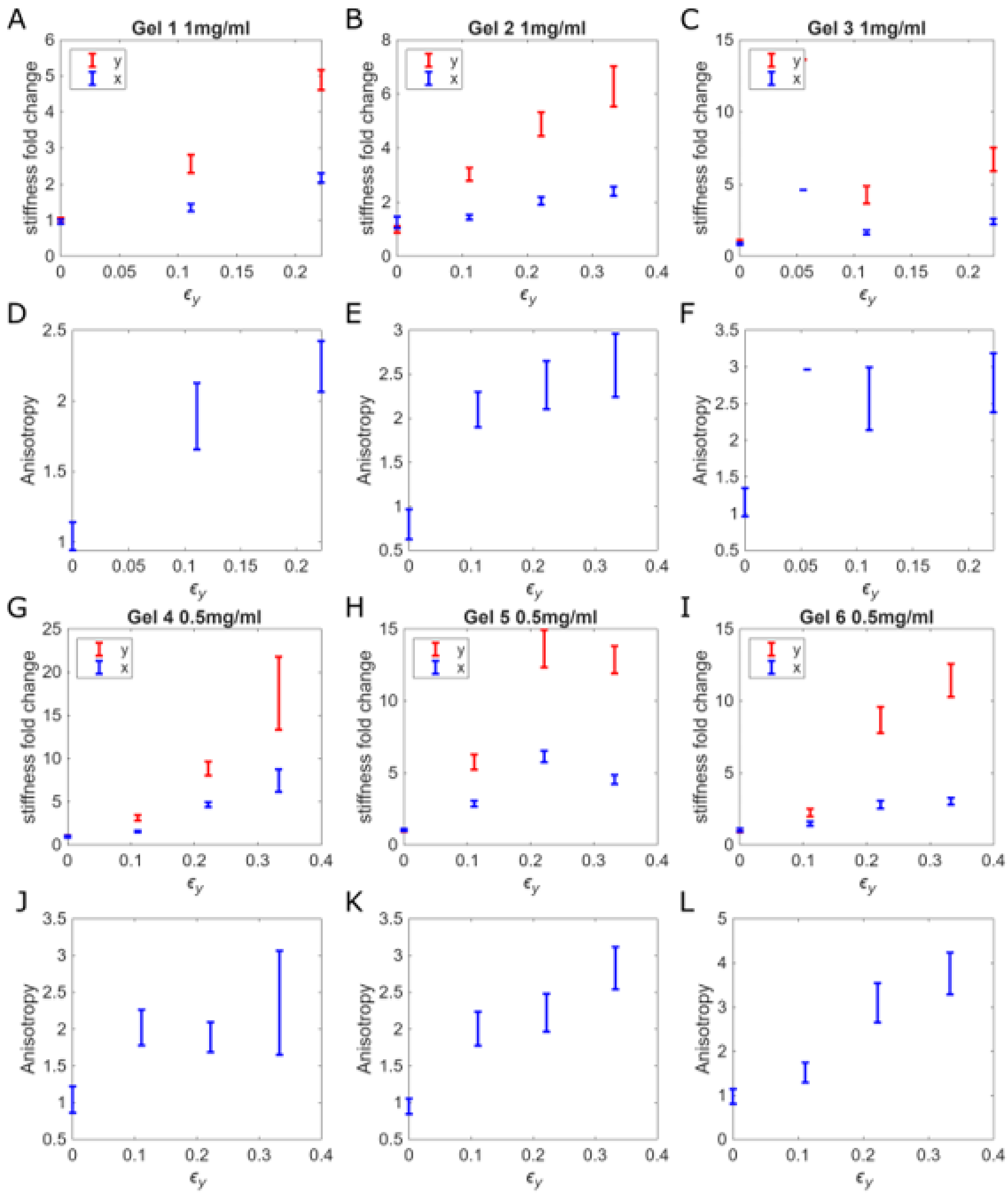

**Fig. S1. Microrheology of stretched fibrin gels - data for individual gels.** Fold-change in the local stiffness in the x and y directions and Y/X anisotropy at 2 Hz with increasing strain, for 0.5 mg/ml fibrin (A-F), and 1 mg/ml fibrin (G-L). Error bars are mean ± standard error for 15 individual beads.

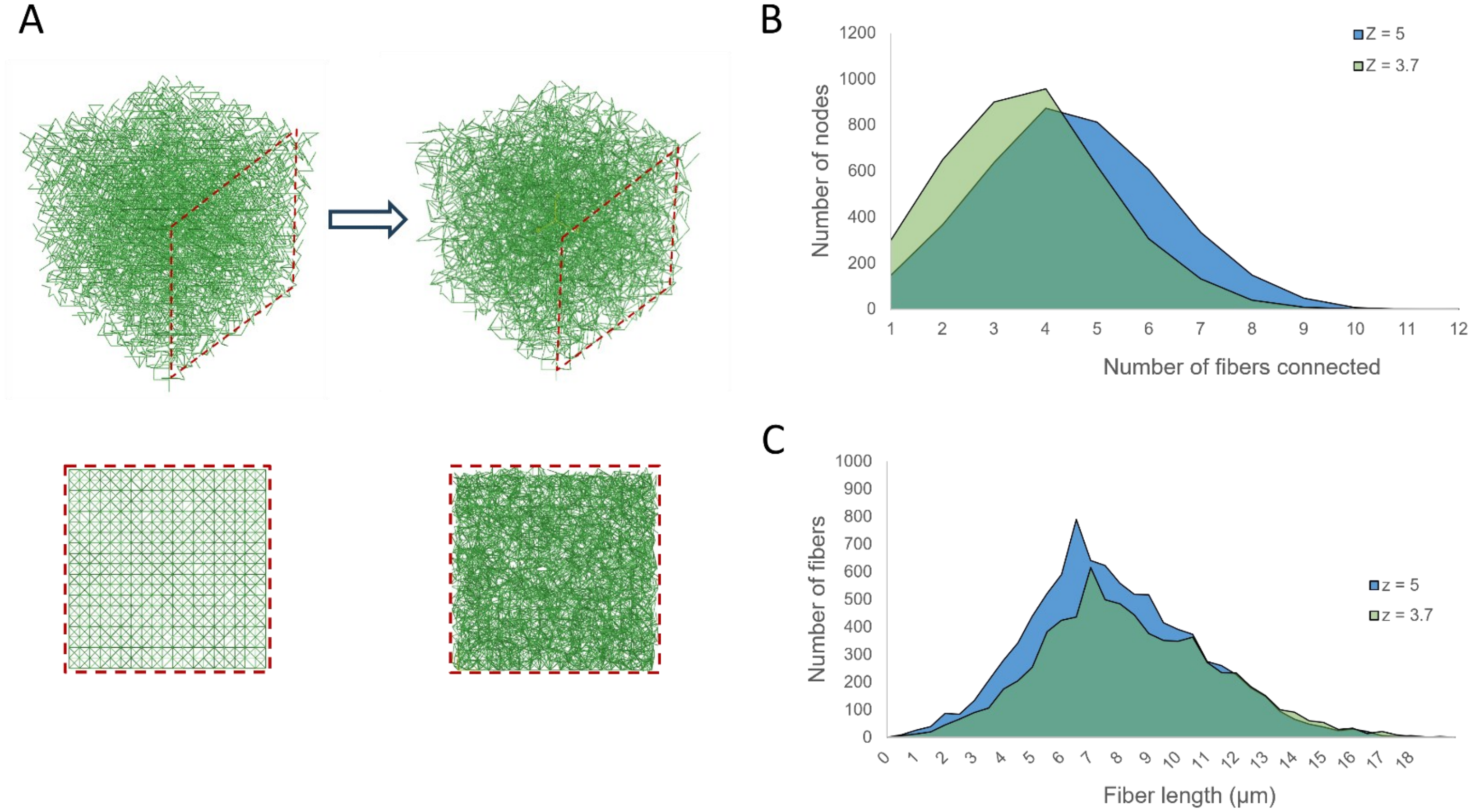

**Fig. S2. Fibrous network model – geometry and characteristics.** (A) The network was initially constructed as an organized lattice of 19×19×19 face-centered-cubic (FCC) repeating units (left). The node positions were randomly adjusted to create an isotropic arrangement of the fibers (right). Side views are shown at the bottom. The network with average connectivity of 3.7 is shown here, but the process of nodal position adjustment was the same for both networks. (B) connectivity distribution in the network. The average connectivity obtained after random elimination of fibers is either 3.7 or 5.0, which are values that characterize biological biopolymer networks. (C) Fiber length distribution in the network, which was achieved through the randomization process at the node locations.

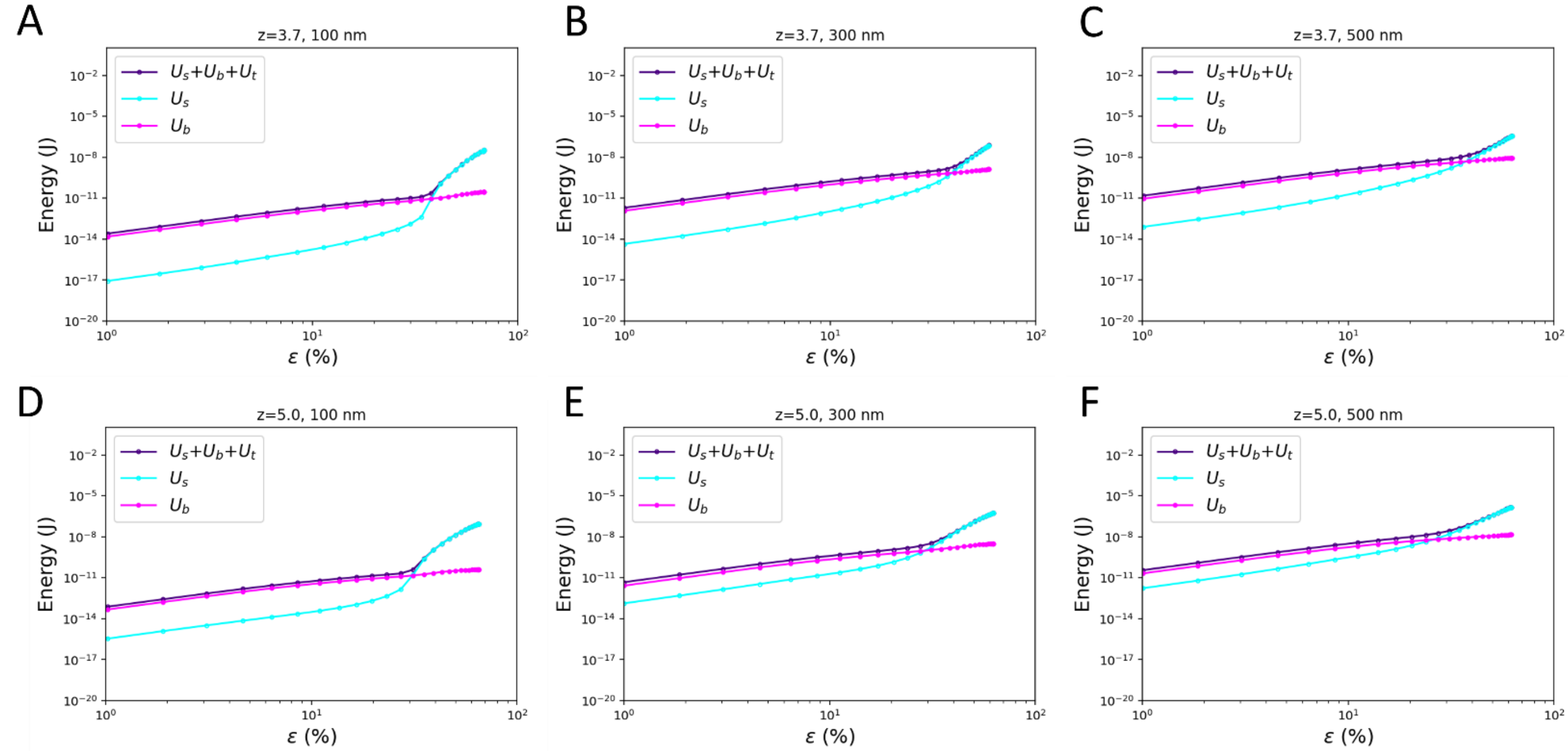

**Fig. S3. Results of uniaxial tensile simulations: full energy analysis.** Energies are shown for the low connectivity networks (A-C) and for high connectivity networks (D-F). While bending and stretching are dominating the deformation in the network, the energy resulting from the torsion of fibers exists, but to a small extent.

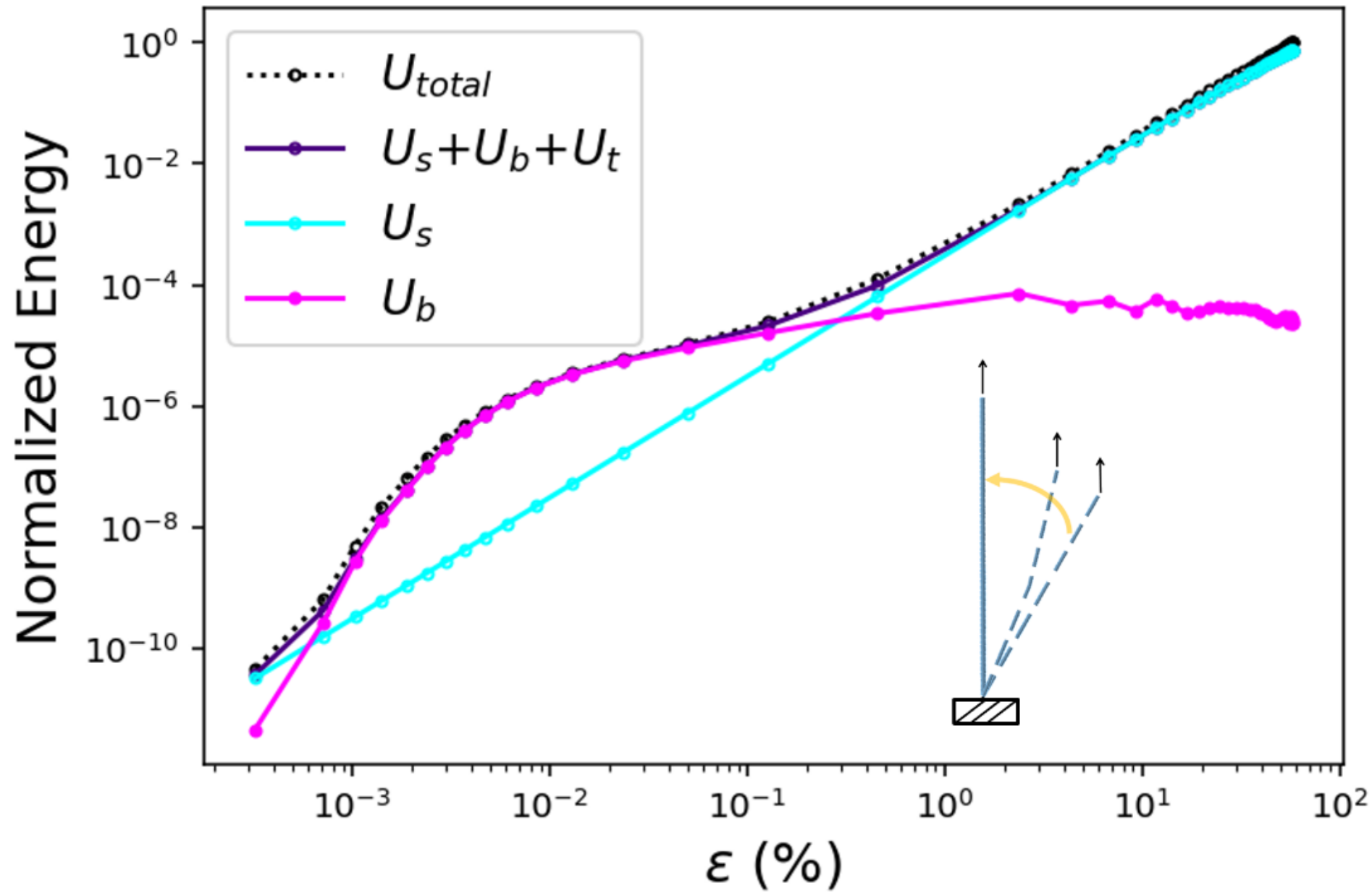

**Fig. S4. Validation for energy calculation: Energy analysis – single fiber (<*z*> = 3.7, *Df* = 300nm).** Response of a single fiber fixed at the bottom and pulled upwards as shown in the inset. At first the deformation is dominated by fiber bending and at ~0.5% strain the fiber is fully aligned vertically and stretching energy starts to dominate the response, as opposed to the full network, where it takes higher global strain for most of the fibers to align in the direction of stretching.

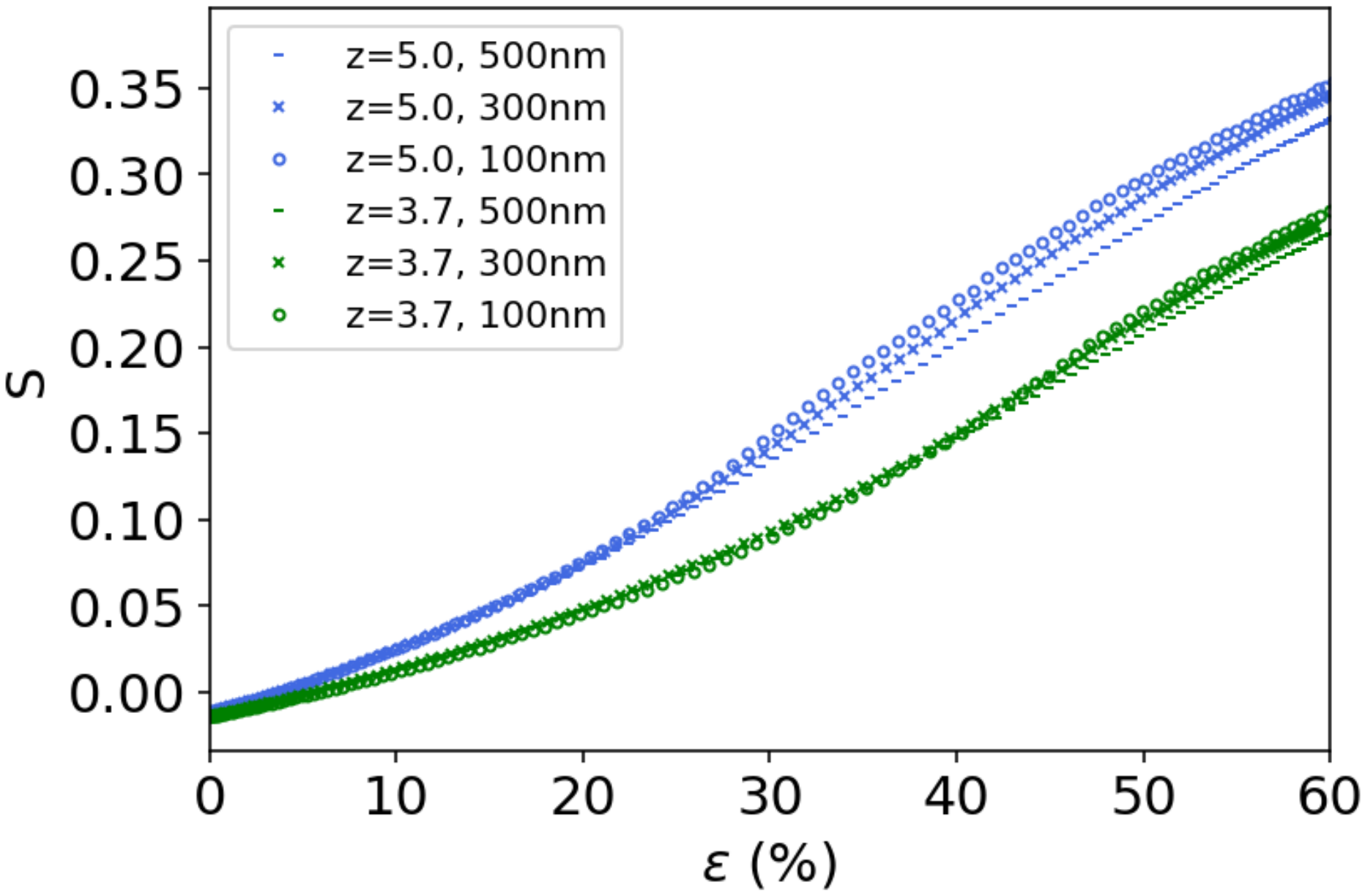

**Fig. S5. The nematic order parameter (i.e., NOP) of the networks vs. strain.** All networks demonstrate alignment of the fibers in the direction of tension (y) as they are stretched, but the alignment is greater in the higher connectivity networks.

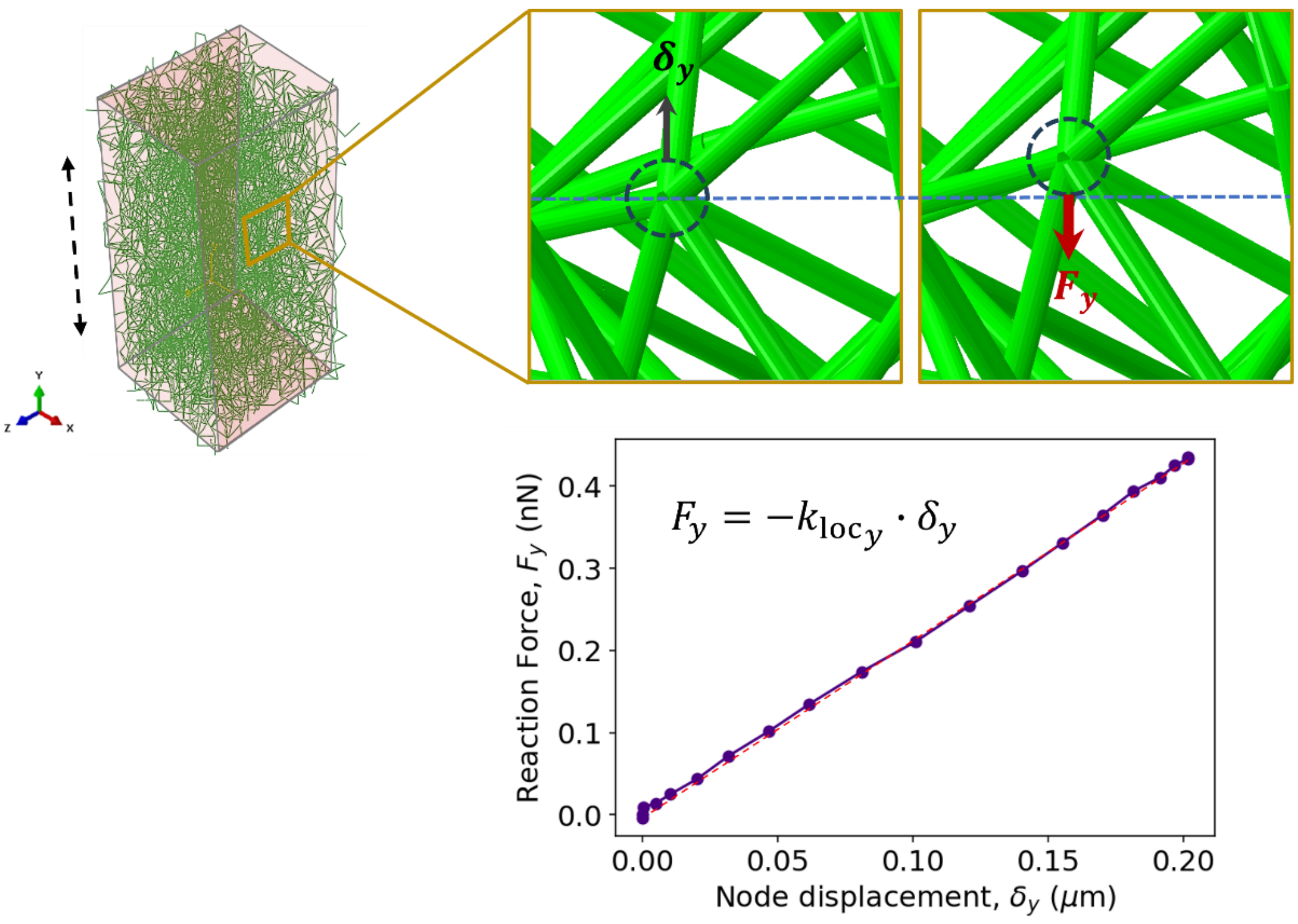

**Fig. S6. Local stiffness calculation.** In a particular level of stretch of the network, a displacement δ=0.2 μm was applied to a node along either the stretching direction (y) or the transverse direction (x), and the resulting reaction force component along that direction was reported. For illustration, the example shown here is for δy on node 2122 (marked with a dashed circle) for the case of ε=35%, *Df* = 300nm, <*z*> = 3.7. Local stiffness was calculated from the force-versus-displacement graph. Since this relation was approximately linear at such a small displacement range, the local stiffness in that direction, $k_{\text{loc}_y}$, was determined as the slope obtained by fitting a straight line to the curve (red dashed line).

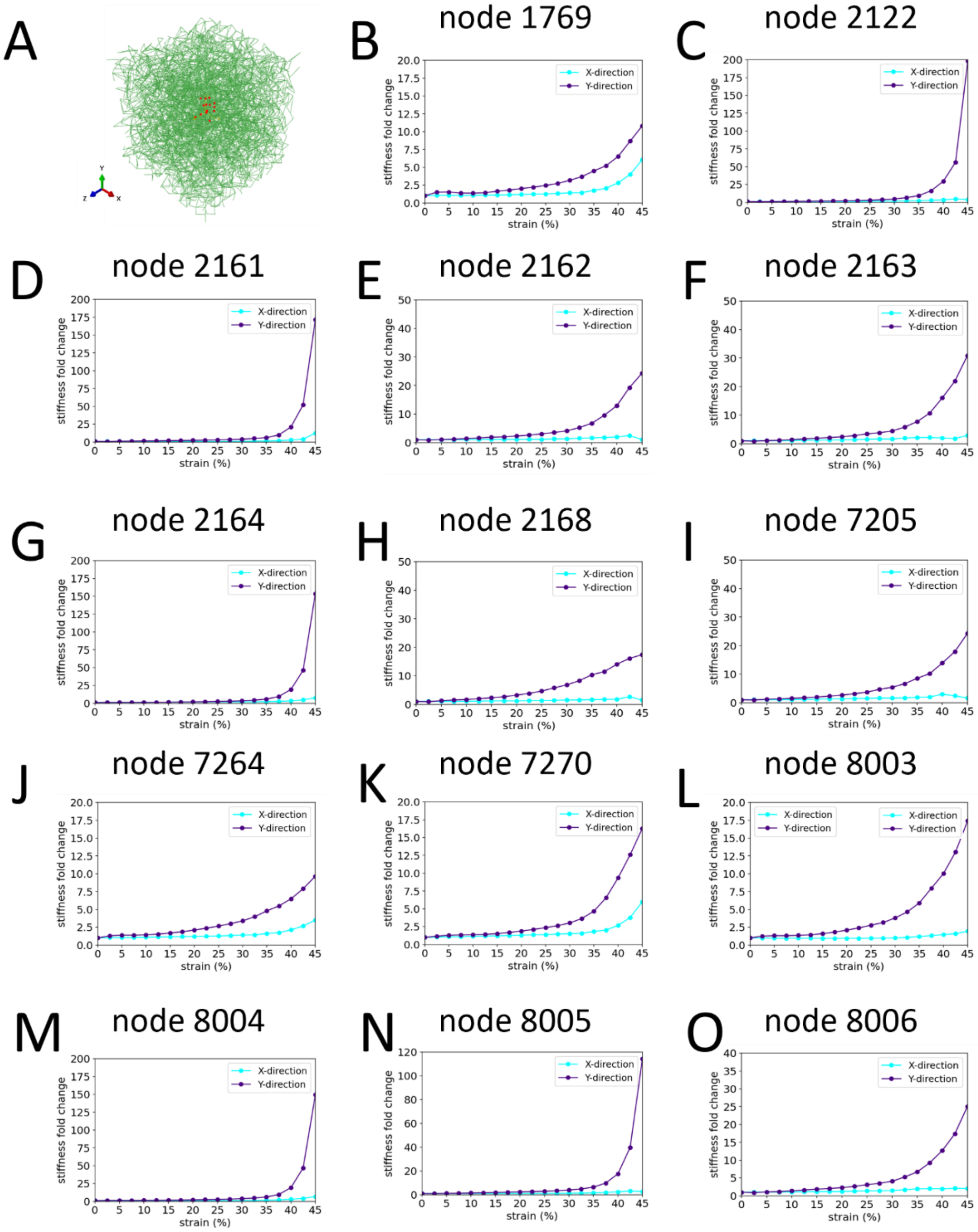

**Fig. S7. Fold-change in the local stiffness in the x and y directions - data of single nodes for $<z>$ = 3.7, *Df* = 100nm.** (A) An image of the network with the nodes in the middle, where the local stiffnesses were measured (marked in red). (B-O) Fold-change of local stiffnesses.

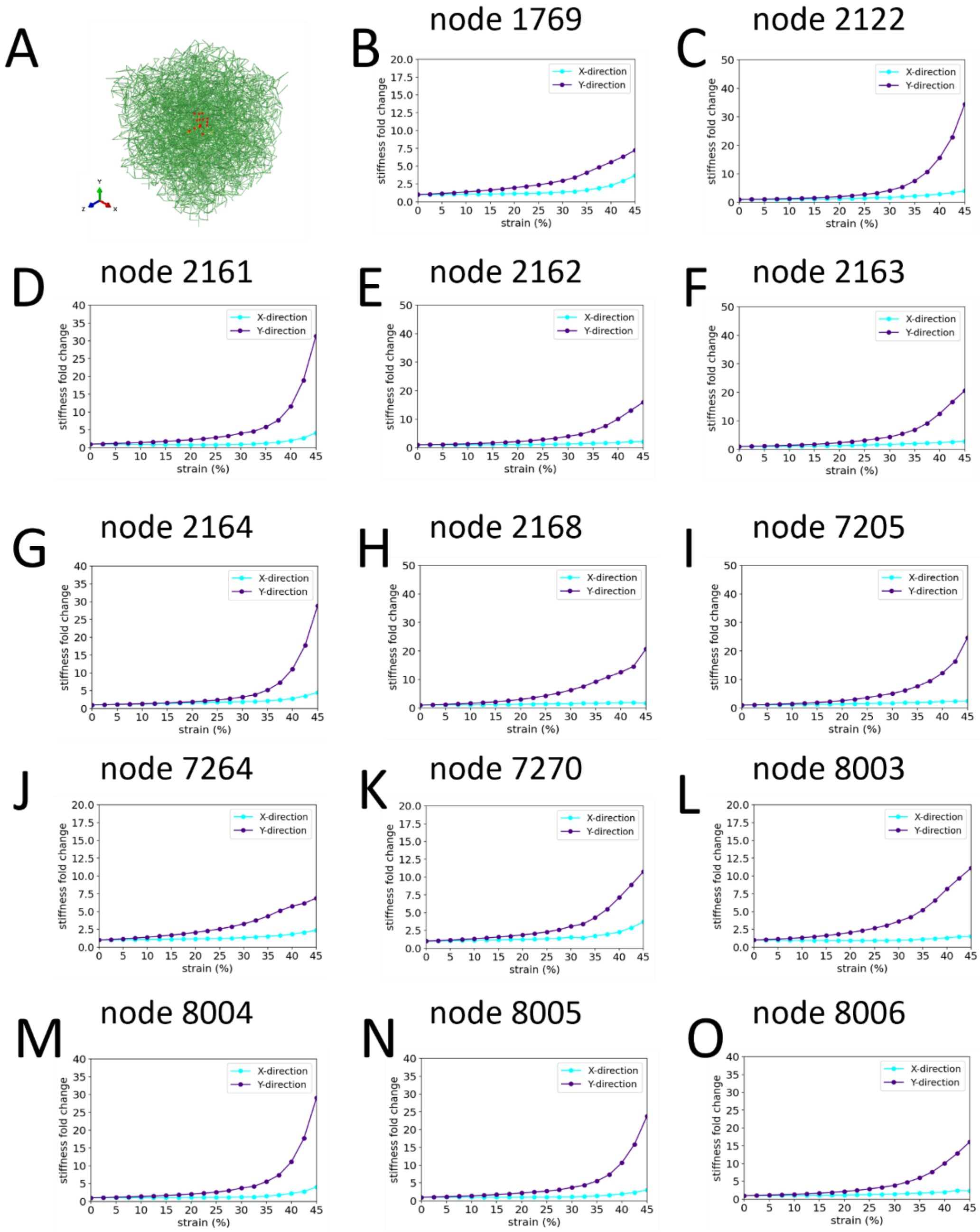

**Fig. S8. Fold-change in the local stiffness in the x and y directions - data of single nodes for $<z>$ = 3.7, $Df$ = 300nm.** (A) An image of the network with the nodes in the middle, where the local stiffnesses were measured (marked in red). (B-O) Fold-change of local stiffnesses.

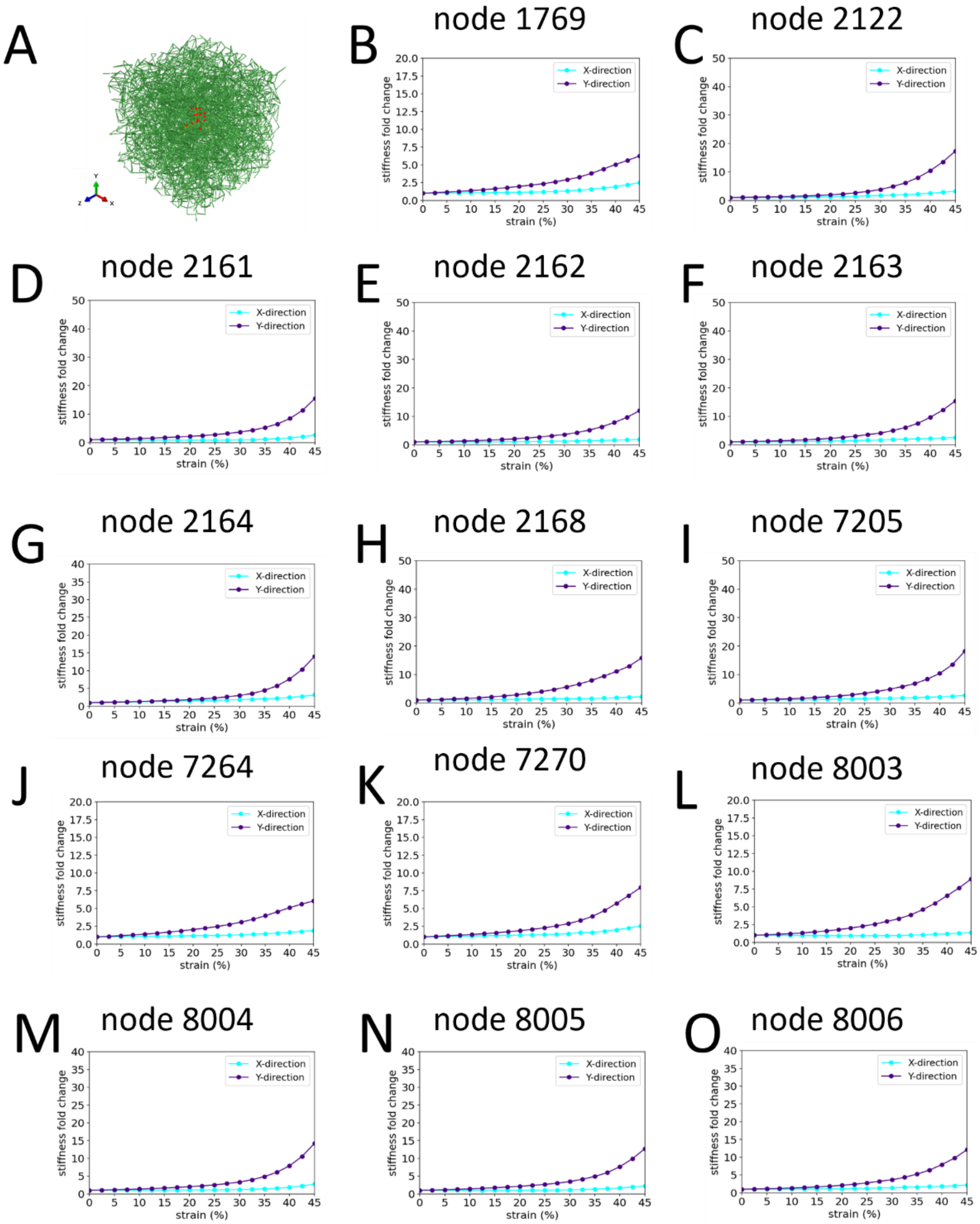

**Fig. S9. Fold-change in the local stiffness in the x and y directions - data of single nodes for <*z*> = 3.7, *Df* = 500nm.** (A) An image of the network with the nodes in the middle, where the local stiffnesses were measured (marked in red). (B-O) Fold-change of local stiffnesses.

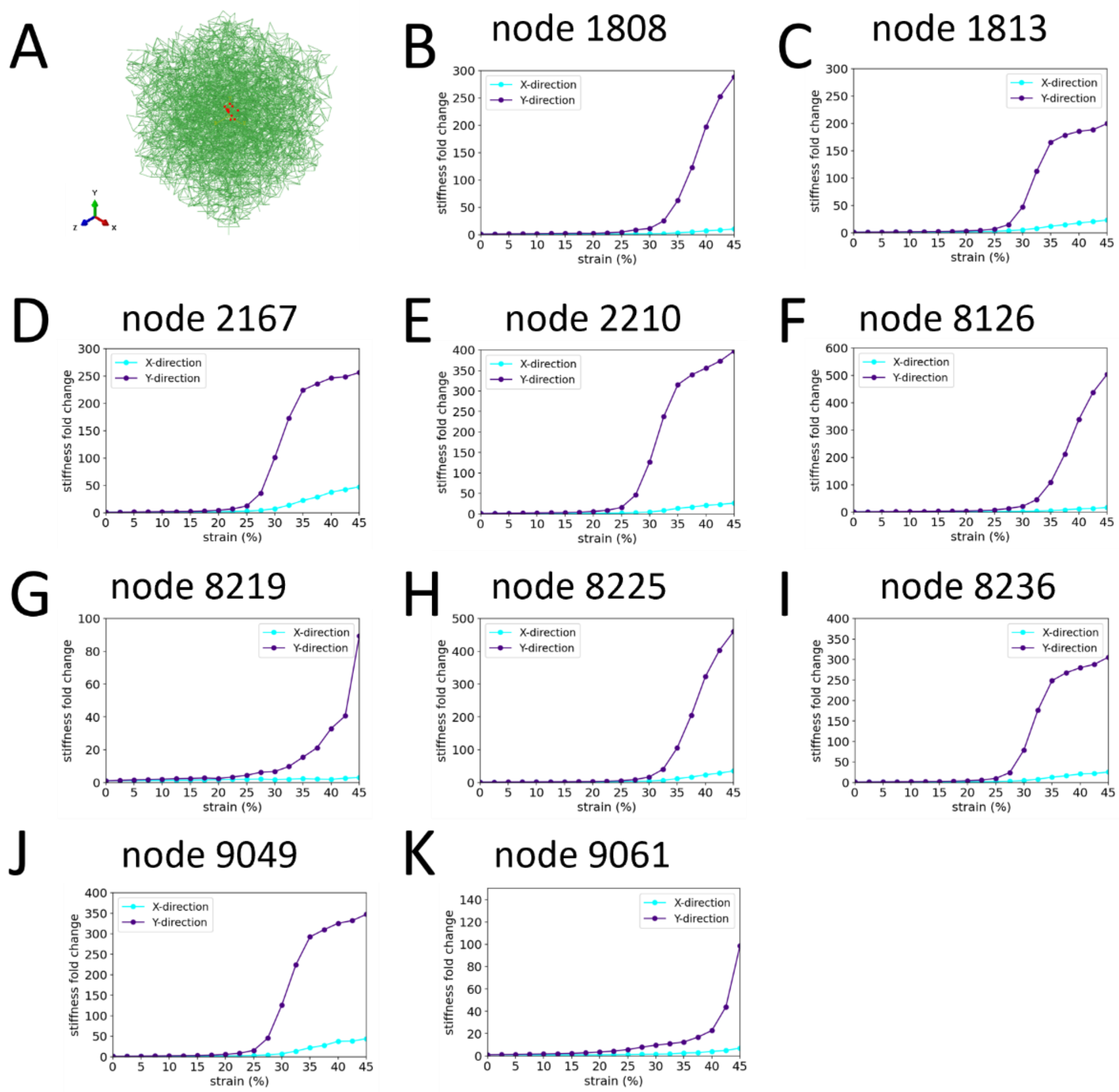

**Fig. S10. Fold-change in the local stiffness in the x and y directions - data of single nodes for <*z*> = 5.0, *Df* = 100nm.** (A) An image of the network with the nodes in the middle, where the local stiffnesses were measured (marked in red). (B-K) Fold-change of local stiffnesses.

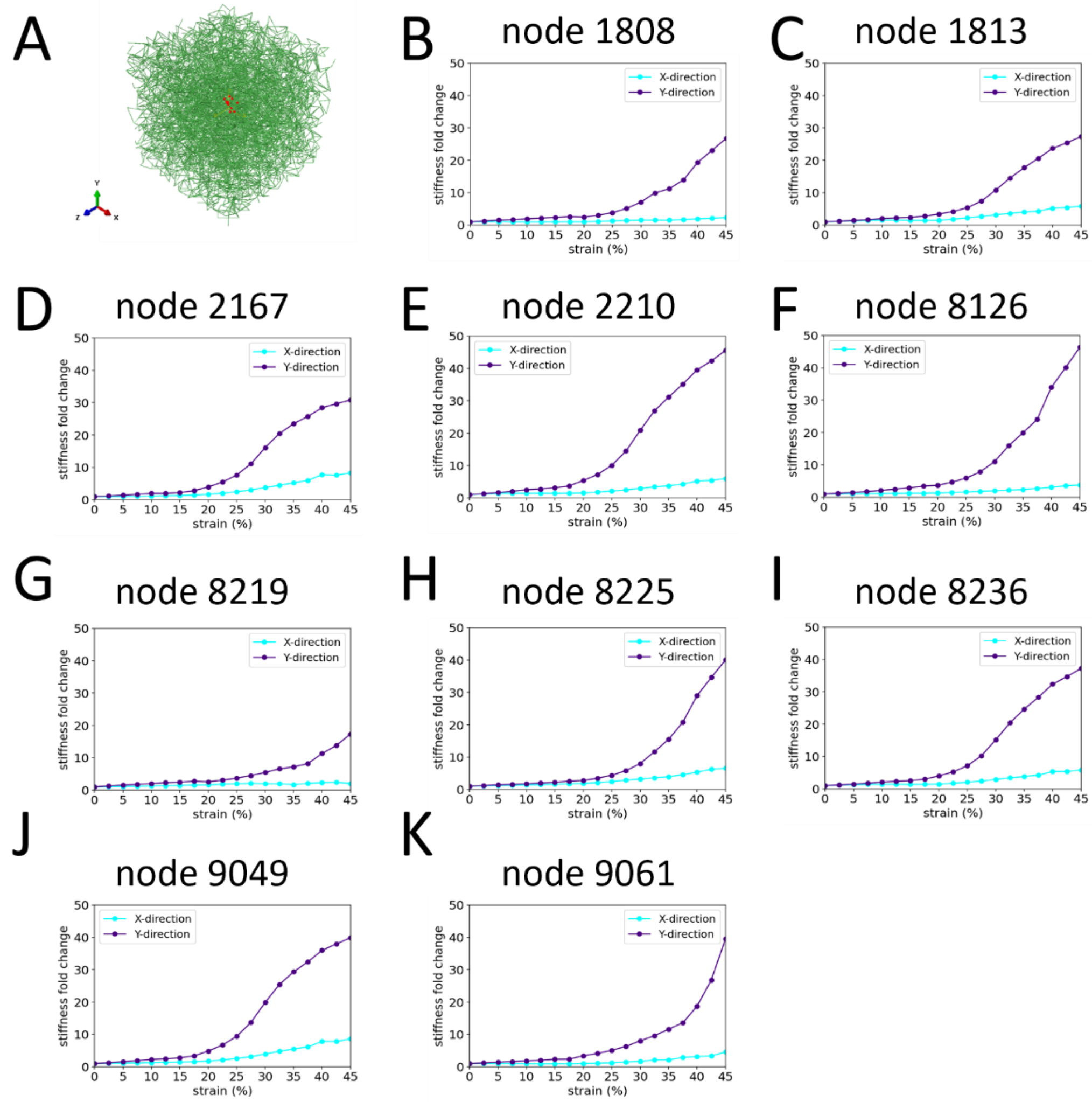

**Fig. S11. Fold-change in the local stiffness in the x and y directions - data of single nodes for <*z*> = 5.0, *Df* = 300nm.** (A) An image of the network with the nodes in the middle, where the local stiffnesses were measured (marked in red). (B-K) Fold-change of local stiffnesses.

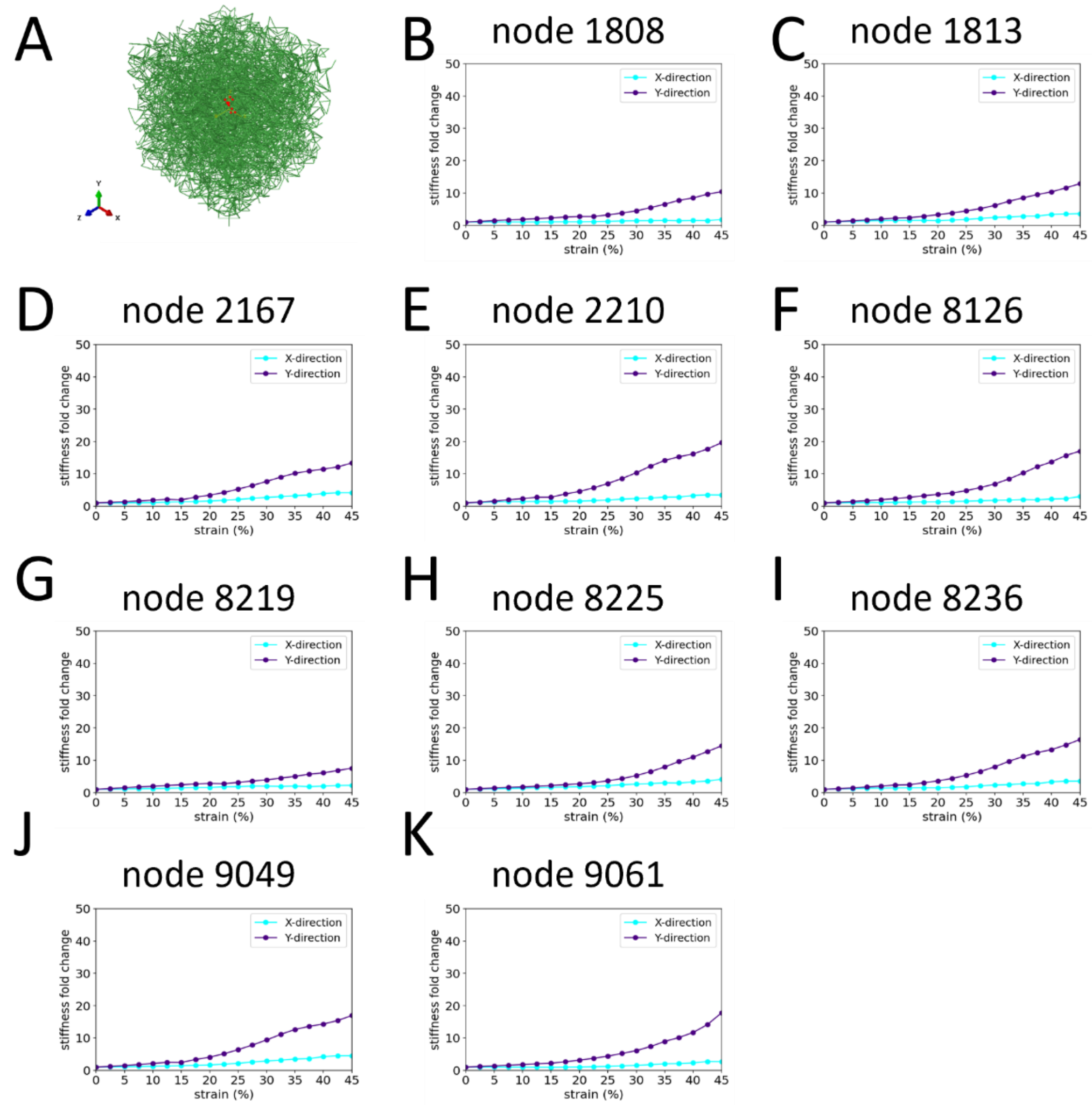

**Fig. S12. Fold-change in the local stiffness in the x and y directions - data of single nodes for <*z*> = 5.0, *Df* = 500nm.** (A) An image of the network with the nodes in the middle, where the local stiffnesses were measured (marked in red). (B-K) Fold-change of local stiffnesses.

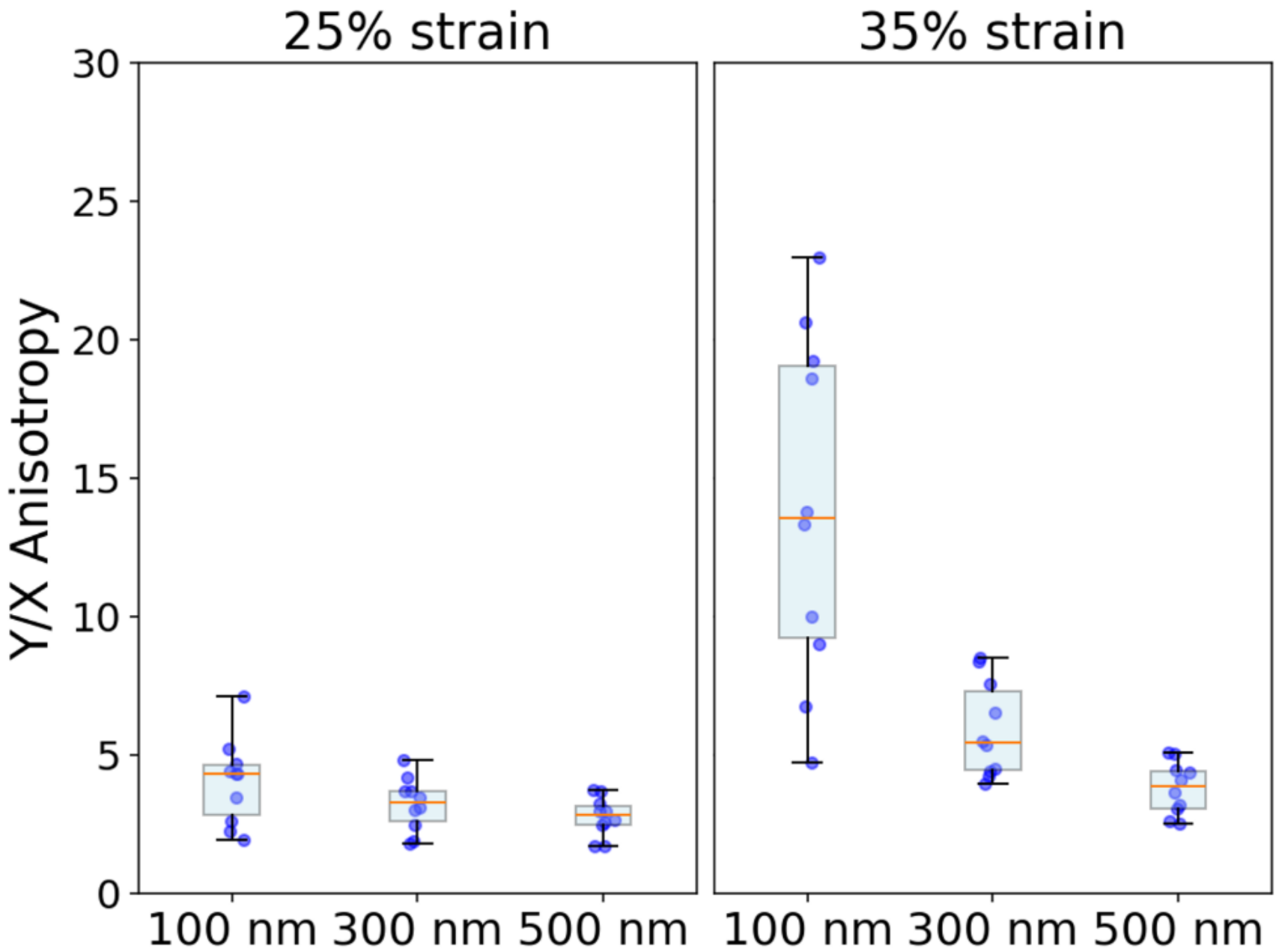

**Fig. S13.** Distributions of the Y/X anisotropy <*z*> = 5.0, measured at the nodes in the middle of the network before and after $\varepsilon_{\mathrm{cr}}$, corresponding to strains of 25% and 35%.